\documentclass[pdftex,usegraphicx,usenatbib]{mn2e}

\usepackage{graphicx}
\usepackage{natbib}
\usepackage{mathptmx}
\usepackage{amsmath}
\usepackage{amssymb}
\usepackage{wasysym}
\usepackage{url}
\usepackage{nohyperref}
\usepackage{times}
\usepackage{epsfig}
\usepackage[T1]{fontenc}
\usepackage[normalem]{ulem}

\newcommand{\apj}{ApJ}
\newcommand{\mnras}{MNRAS}
\newcommand{\nat}{Nat}

\newcommand{\aap}{A\&A}                   
\newcommand{\apjs}{ApJS}                  
\newcommand{\apjl}{ApJ}                   
\newcommand{\pasj}{PASJ}
\newcommand{\gca}{Geochim. Cosmochim. Acta}
\newcommand{\physrep}{Phys. Rep.}
\newcommand{\aapr}{A\&A Rev.}
\newcommand{\ssr}{Space Sci. Rev.}

\newcommand{\jgr}{J. Geophys. Res.}
\newcommand{\prl}{PhRvL}

\newcommand{\Mpc}{\rm\; Mpc}
\newcommand{\kpc}{\rm\; kpc}

\newcommand{\km}{\rm\; km}

\newcommand{\cm}{\rm\; cm}

\newcommand{\pix}{\rm\; pixel}

\newcommand{\ppix}{\hbox{$\pix^{-1}\,$}}

\newcommand{\yr}{\rm\; yr}
\newcommand{\Gyr}{\rm\; Gyr}

\newcommand{\s}{\rm\; s}

\newcommand{\Ms}{\rm\; Ms}

\newcommand{\K}{\rm\; K}

\newcommand{\pseudopressure}{\hbox{$\keV\cm^{-5/2}\pasecsq\,$}}

\newcommand{\keV}{\rm\; keV}
\newcommand{\eV}{\rm\; eV}
\newcommand{\erg}{\rm\; erg}

\newcommand{\expmapcorr}{\hbox{$\rm\thinspace counts\pcmsq\ps\ppix$}}
\newcommand{\kmps}{\hbox{$\km\s^{-1}\,$}}

\newcommand{\kmpspMpc}{\hbox{$\kmps\Mpc^{-1}\,$}}

\newcommand{\Zsun}{\hbox{$\thinspace \mathrm{Z}_{\odot}$}}

\newcommand{\amin}{\rm\; arcmin}

\newcommand{\asec}{\rm\; arcsec}

\newcommand{\pcmsq}{\hbox{$\cm^{-2}\,$}}
\newcommand{\pcmcu}{\hbox{$\cm^{-3}\,$}}

\newcommand{\ps}{\hbox{$\s^{-1}\,$}}

\newcommand{\pasecsq}{\hbox{$\asec^{-2}\,$}}

\begin{document}

\title[The structure of cluster merger shocks]{The structure of cluster merger shocks: turbulent width and the electron heating timescale}
\author[H.R. Russell et al.]  
    {\parbox[]{7.in}{H.~R. Russell$^{1}$\thanks{E-mail: 
          helen.russell@nottingham.ac.uk},
        P.~E.~J. Nulsen$^{2,3}$,
        D. Caprioli$^{4,5}$,
        U. Chadayammuri$^{2}$,
        A.~C. Fabian$^6$,
        M.~W. Kunz$^{7,8}$,
        B.~R. McNamara$^{9,10}$,
        J.~S. Sanders$^{11}$,
        A. Richard-Laferri\`ere$^{6}$,
        M. Beleznay$^{12}$,
        R.~E.~A. Canning$^{13}$,
        J. Hlavacek-Larrondo$^{14}$ and
        L.~J. King$^{15}$
        \\   
     \footnotesize
     $^1$ School of Physics \& Astronomy, University of Nottingham, University Park, Nottingham NG7 2RD, UK\\  
     $^2$ Center for Astrophysics | Harvard \& Smithsonian, 60 Garden Street, Cambridge, MA 02138, USA\\
     $^3$ ICRAR, University of Western Australia, 35 Stirling Hwy, Crawley, WA 6009, Australia\\
     $^4$ Department of Astronomy and Astrophysics, University of Chicago, 5640 S Ellis Ave, Chicago, IL 60637, USA\\
     $^5$ Enrico Fermi Institute, University of Chicago, 5640 S Ellis Ave, Chicago, IL 60637, USA\\
     $^6$ Institute of Astronomy, Madingley Road, Cambridge CB3 0HA, UK\\
     $^7$ Department of Astrophysical Sciences, University of Princeton, 4 Ivy Ln, Princeton, NJ 08544, USA\\
     $^8$ Princeton Plasma Physics Laboratory, PO Box 451, Princeton, NJ 08543, USA\\
     $^{9}$ Department of Physics and Astronomy, University of Waterloo, Waterloo, ON N2L 3G1, Canada\\
     $^{10}$ Perimeter Institute for Theoretical Physics, Waterloo, Canada\\
     $^{11}$ Max-Planck-Institut f{\"u}r extraterrestrische Physik, Gie{\ss}enbachstra{\ss}e 1, D-85748, Garching, Germany\\   
     $^{12}$ Kavli Institute for Astrophysics and Space Research, Massachusetts Institute of Technology, 77 Massachusetts Avenue, Cambridge, MA 02139, USA\\
     $^{13}$ Institute of Cosmology and Gravitation, University of Portsmouth, Portsmouth, PO1 3FX, UK\\
     $^{14}$ D{\'e}partement de Physique, Universit{\'e} de Montr{\'e}al, Succ. Centre-Ville, Montr{\'e}al, H3C 3J7, Canada\\
     $^{15}$ Department of Physics, University of Texas at Dallas, 800 W Campbell Rd, Richardson, TX 75080, USA\\
  }
    }
   
\maketitle

\begin{abstract}
  We present a new $2\Ms$ \textit{Chandra} observation of the cluster
  merger Abell\,2146, which hosts two huge $M{\sim}2$ shock fronts
  each ${\sim}500\kpc$ across.  For the first time, we resolve and
  measure the width of cluster merger shocks.  The best-fit width for
  the bow shock is $17\pm1\kpc$ and for the upstream shock is
  $10.7\pm0.3\kpc$.  A narrow collisionless shock will appear broader
  in projection if its smooth shape is warped by local gas motions.
  We show that both shock widths are consistent with collisionless shocks blurred by local gas motions of $290\pm30\kmps$.  The upstream shock forms later on in
  the merger than the bow shock and is therefore expected to be
  significantly narrower.  From the electron temperature profile
  behind the bow shock, we measure the timescale for the electrons and
  ions to come back into thermal equilibrium.  We rule out rapid
  thermal equilibration of the electrons with the shock-heated ions at
  the $6\sigma$ level.  The observed temperature profile instead
  favours collisional equilibration.  For these cluster merger shocks, which have low sonic Mach numbers and propagate through a high $\beta$ plasma, we find no evidence for electron heating over that produced by adiabatic compression.  Our findings are expected to be valid for collisionless shocks with similar parameters in other environments and support the existing picture from the solar wind and supernova remnants.  The upstream shock is consistent with this
  result but has a more complex structure, including a $\sim2\keV$
  increase in temperature ${\sim}50\kpc$ ahead of the shock.
\end{abstract}


\section{Introduction}
\label{sec:intro}

\begin{figure*}
  \begin{minipage}{\textwidth}
    \centering
    \includegraphics[width=0.98\columnwidth]{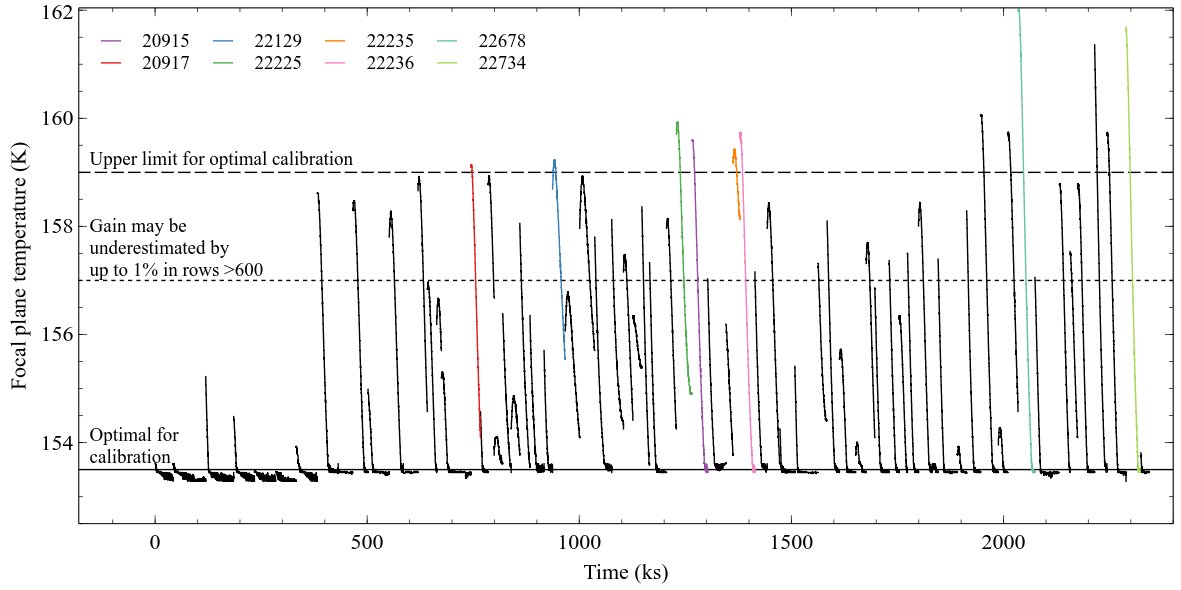}
    \caption{Focal-plane temperature over time for each observation, where the observation start times have been modified to continue from the end of the previous observation.  The first 8 observations were taken from August to October 2010 when \textit{Chandra}'s thermal performance was more stable.  Observations which were noted in validation and verification to have periods of higher than optimal focal-plane temperature are shown by the coloured lines.}
    \label{fig:fptemp}
  \end{minipage}
\end{figure*}

\begin{figure}
  \centering
  \includegraphics[width=0.98\columnwidth]{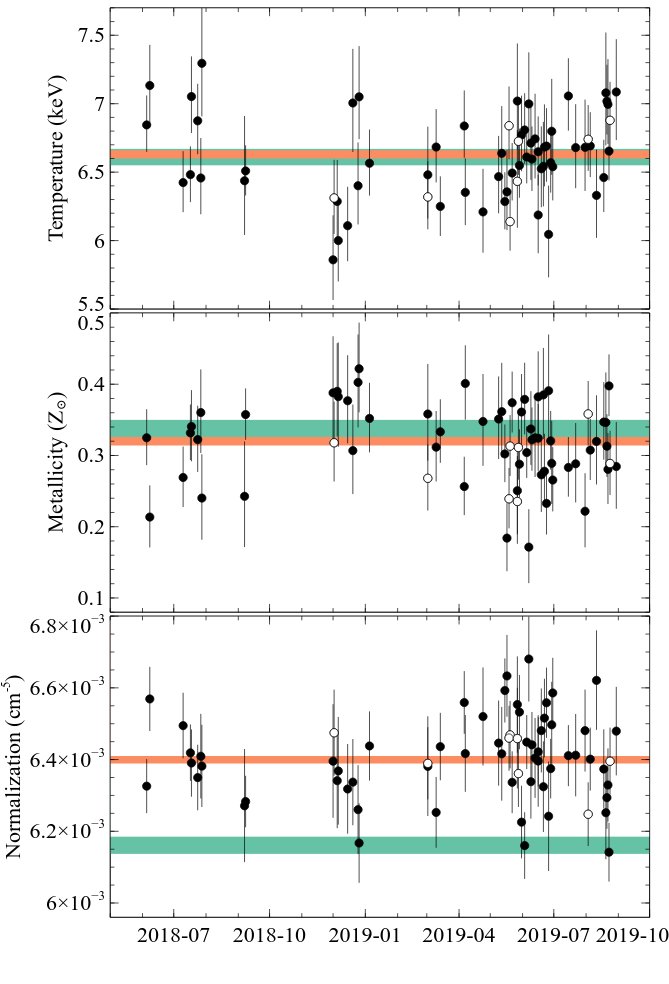}
  \caption{Comparison of the best-fit temperature, metallicity and normalization for a spectrum extracted from a 1.5 arcmin radius region covering most of the cluster emission in each dataset.  The green and orange coloured regions show the 1$\sigma$ uncertainties for each parameter from fitting only the 2010 or the new observations, respectively.  The white points show the results for the observations that were flagged as having periods of higher than optimum focal-plane temperature.}
  \label{fig:fppars}
\end{figure}

Major mergers of massive galaxy clusters are the primary hierarchical
growth mechanism for clusters (for a review see
\citealt{Markevitch07}).  During a merger, the galaxies and dark
matter in the clusters behave as nearly collisionless particles
(e.g. \citealt{Clowe06}) and move unimpeded ahead of the spectacular
clash between the hot intracluster atmospheres.  X-ray observations of
the atmospheric plasma reveal shock fronts, sharp edges associated
with cold fronts, large-scale turbulent eddies and tails of gas
stripped by ram pressure (e.g. \citealt{Markevitch06}).  Shocks and
turbulence generated by the merger accelerate particles to
relativistic speeds.  The resulting synchrotron emission produces
large-scale radio halos and relics (for a review see
\citealt{vanWeeren19}).

The shocks, with typical Mach numbers $M\sim1-3$, dissipate most of
the merger's $\sim10^{64}\erg$ of kinetic energy
(e.g. \citealt{Sarazin02}).  At low atmospheric densities, shock
fronts are believed to be collisionless.  The kinetic energy of the
inflowing gas is dissipated via plasma-wave interactions between the
particles and the magnetic field (e.g. \citealt{Treumann09}).
Spacecraft travelling through the collisionless solar-wind shock have
revealed ions are heated in a narrow shock layer whose width is of order
their Larmor radius (e.g. \citealt{Schwartz88}).  Electrons may remain
significantly cooler, equilibrating with the ions via collisions,
unless there is an additional electron heating process (for a review
see \citealt{Ghavamian13}).  \textit{Chandra} observations of merger
shocks can map the postshock electron temperature (for a review see
e.g. \citealt{Bohringer10}), determine the Mach number and shock speed
and thereby measure the electron heating timescale in a single
observation (e.g. \citealt{Markevitch06}).

Detecting a shock front with a sharp density edge and an unambiguous
jump in temperature is rare.  Only a handful are known
(e.g. \citealt{Markevitch07}).  This is primarily due to an observational
shortcoming against detecting well-defined shock fronts.  Shock fronts are
easiest to detect shortly after the first pericentre passage and if
the merger axis is oriented close to the plane of the sky.  Only three clusters host shock fronts bright
enough for detailed study: the Bullet cluster
(\citealt{Markevitch06}), Abell\,520 (\citealt{Markevitch05};
\citealt{Wang18}) and Abell\,2146.  Abell\,2146 hosts two $M{\sim}2$ merger
shock fronts propagating in opposite directions
(\citealt{Russell10,Russell12}).  Although the Bullet cluster shock
front has the highest Mach number, the gas temperature (${>}30\keV$)
is difficult to constrain with Chandra's lower energy range.  Abell\,520
hosts a weaker shock front.  Its postshock gas is permeated with
substructure related to the disintegrating subcluster cool core, which
must be carefully masked out (\citealt{Wang18}).  Abell\,2146 hosts
the two brightest and cleanest shock fronts at measurable temperatures
and was therefore the recent target of a \textit{Chandra} legacy-class
observation of a cluster merger.

Here we present the new $2\Ms$ \textit{Chandra} observation of Abell
2146.  We discuss new structures revealed in this deep dataset and
focus on the detailed structure of the shock fronts and on measuring
the electron-ion thermal equilibration timescale.  Detailed analyses
of the break up of the cool cores, constraints on the rate of
conduction and level of turbulence will be covered in separate papers.
We assume $H_0=70\kmpspMpc$, $\Omega_{\mathrm{m}}=0.3$ and
$\Omega_\Lambda=0.7$, translating to a scale of $3.7\kpc$ per arcsec
at the redshift $z=0.234$ of Abell\,2146 (\citealt{Struble99};
\citealt{Bohringer00}).  All errors are $1\sigma$ unless otherwise
noted.

\section{Data reduction}
\label{sec:data}

The new \textit{Chandra} observation of the cluster merger Abell\,2146 has an exposure time of $1.93\Ms$ split over 67 separate observations on the ACIS-I detector between June 2018 and August 2019.  When combined with the earlier ACIS-I observations taken in 2010 (\citealt{Russell12}), the total exposure is $2.31\Ms$ (Table \ref{tab:obs}).  All 75 data sets were reprocessed following standard reduction procedures using \textsc{ciao} v4.13 and \textsc{caldb} v4.9.4 provided by the \textit{Chandra} X-ray Center.  These include the latest calibration measurements and crucial updates to the ACIS contaminant model.  Improved background screening provided by VFAINT mode was applied to all observations.  Background light curves were extracted from neighbouring CCDs and filtered using the \textsc{lc\_clean} script to remove periods affected by flares.  The net exposure times are given in Table \ref{tab:obs}.  Only obs. ids 20921 and 21674 were affected significantly by flares.

Abell\,2146 is located at high ecliptic latitude (declination $+66\deg$) and is therefore at a thermally unfavourable pitch angle for \textit{Chandra}, which has an ageing exterior thermal finish.  This deep image was divided into short observations over a long time period.  Several datasets were taken when the focal-plane temperature exceeded the upper limit for optimum calibration of the ACIS gain and spectral resolution.  Fig.~\ref{fig:fptemp} shows the focal-plane temperature over time.  The contrast between the observations taken in 2010, when \textit{Chandra}'s thermal performance was more stable, and the new observations is clear.

Whilst the majority of the observing time was conducted with the optimal focal-plane temperature, we compared the spectral results from different datasets to search for any systematic effect.  Fig.~\ref{fig:fppars} shows the best-fit temperature, metallicity and normalization for the new observations (for an absorbed \textsc{apec} model, see section \ref{sec:spec}).  All are consistent, given the uncertainties, with the equivalent results from the 2010 observations.  No systematic differences are found for the observations with higher than optimum focal-plane temperature.  We compared the best-fit parameters for the new and old observations from 2010 for key regions, such as the profiles across the shock fronts.  The temperature and metallicity values were consistent within the uncertainties (Fig.~\ref{fig:fppars}).  Normalizations appear on average a few per cent higher for the new datasets.  Electron densities derived from these will be ${\sim}2\%$ higher.  This small systematic uncertainty is likely due to the escalating contaminant correction but has minimal impact on our results.  The new datasets are also likely to be affected by small gain changes (see e.g. \citealt{Sanders14}).  We conclude that the calibration is sufficient for our analysis.

The absolute astrometry of each dataset was corrected by cross-matching point sources across the separate observations.  The final event files were reprojected to match the position of obs. ID 21733.  Exposure maps were generated for each observation using energy weights determined from an absorbed \textsc{apec} model at the global cluster temperature of $6.67\pm0.03\keV$, metallicity $0.319\pm0.006\Zsun$ (relative to solar abundances defined by \citealt{AndersGrevesse89} for comparison with previous results) and redshift 0.234.  The absorption was fixed to the Galactic value $n_{\mathrm{H}}=3.0\times10^{20}\pcmsq$ (\citealt{Kalberla05}).


Blank sky backgrounds were generated for each observation.  Each processed identically, reprojected to the corresponding sky position, and normalized to match the count rate in the $9.5$--$12\keV$ energy band.  Following \citet{Vikhlinin05}, we tested additional emission models to account for residual background emission after the blank sky backgrounds were subtracted.  Soft X-ray background residuals due to Galactic foreground were modelled by a $0.18\keV$ \textsc{apec} component with solar metallicity.  Residual unresolved cosmic X-ray background (CXB) was modelled with an absorbed powerlaw with $\Gamma=1.5$.  The best-fit background model parameters were determined for each observation using a spectrum extracted from a large source-free region of the chip.  The normalizations of the soft and unresolved CXB components were consistent with zero within the uncertainties for all observations.  We therefore proceeded with the blank sky background spectra without using additional spectral models.



\section{Image analysis}
\label{sec:image}


\begin{figure*}
  \begin{minipage}{\textwidth}
    \centering
    \includegraphics[width=0.48\columnwidth]{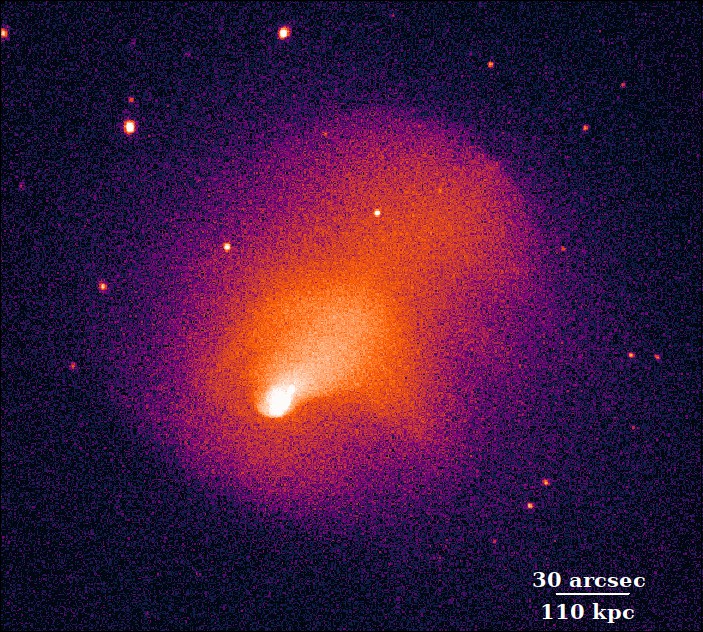}
    \includegraphics[width=0.48\columnwidth]{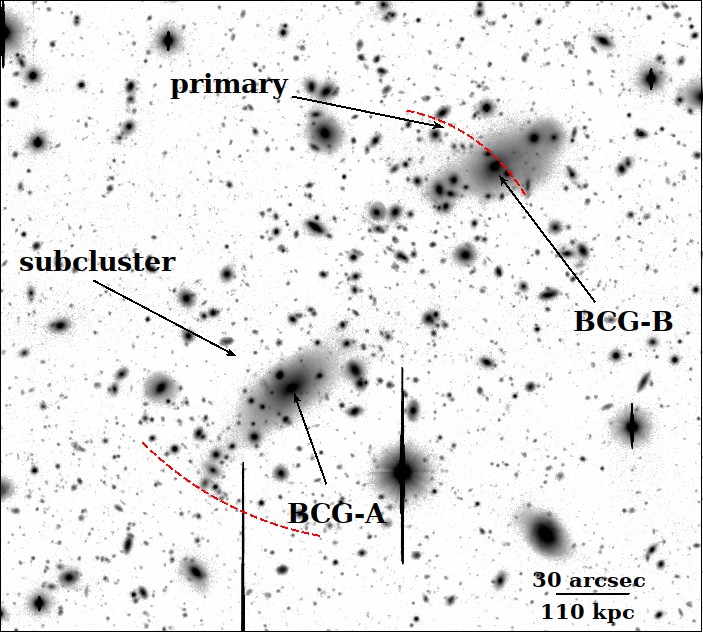}
    \includegraphics[width=0.48\columnwidth]{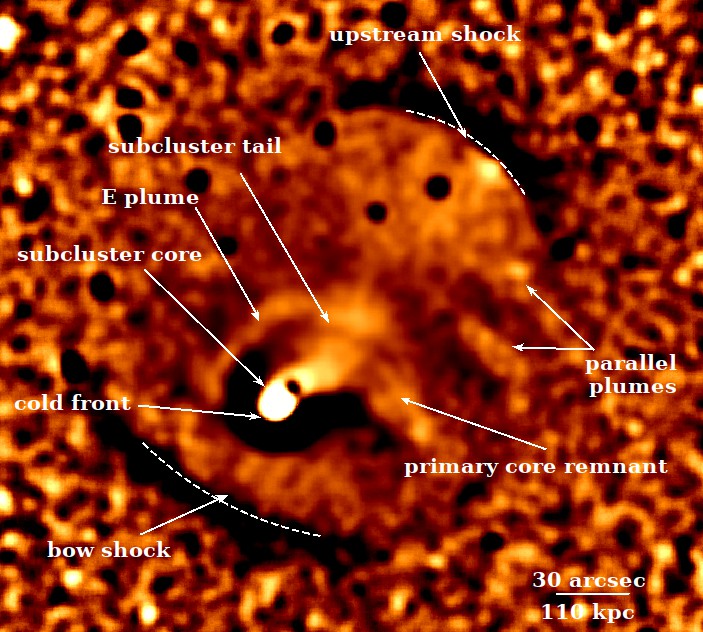}
    \includegraphics[width=0.48\columnwidth]{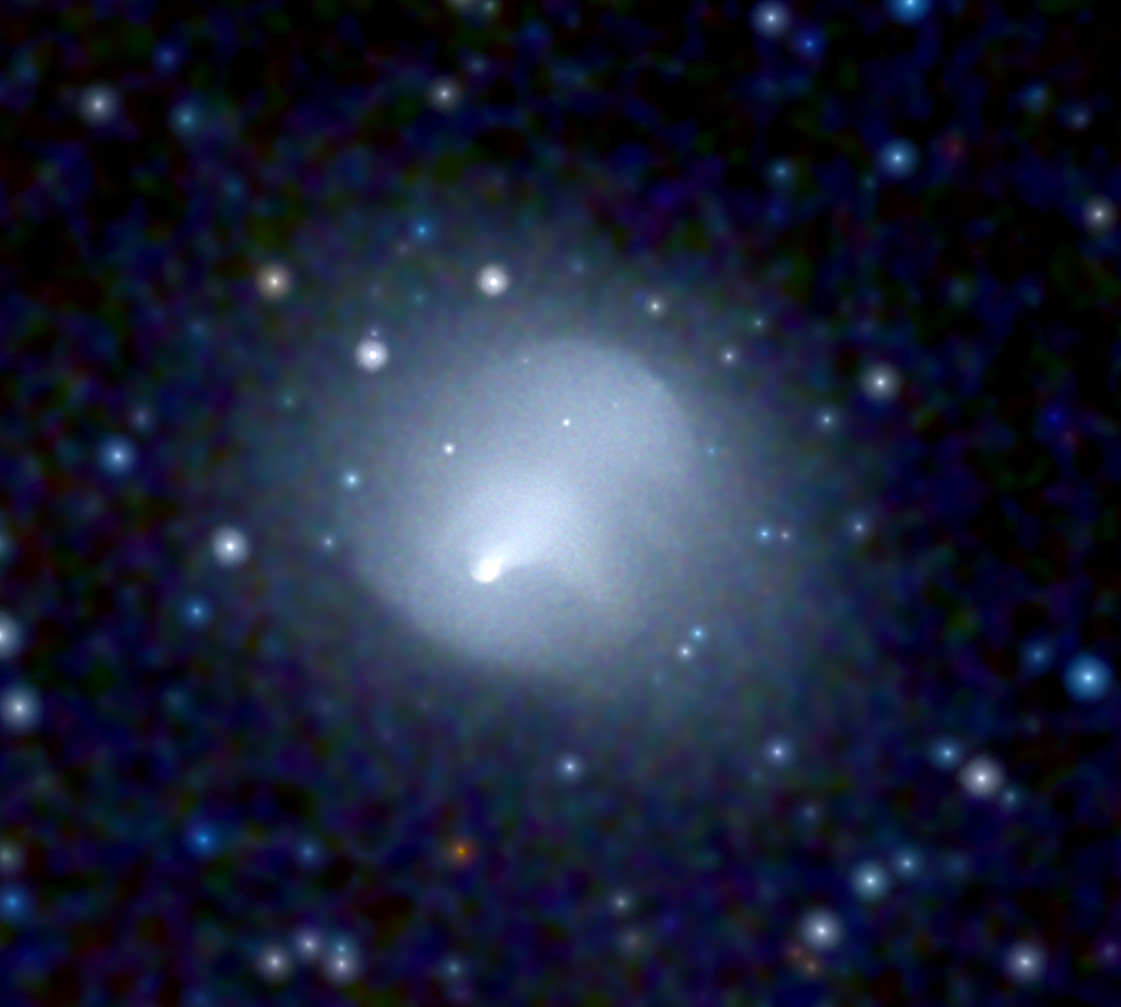}
    \caption{Upper left: Exposure-corrected X-ray image of Abell\,2146 in the $0.5$--$4\keV$ energy band ($\expmapcorr$).  Upper right: Subaru R-band optical image showing the primary and subcluster galaxies (\citealt{King16}).  Lower left: Unsharp-masked image highlighting structure.  The shock front locations determined from surface brightness profile fitting are shown by the arcs.  The exposure-corrected image, optical image and unsharp-masked image cover the same field of view.  Lower right: An RGB image covering a larger field of view with red for $0.5$--$1.5\keV$, green for $1.5$--$2.5\keV$ and blue for $2.5$--$6\keV$.  An adaptive smoothing method has been applied to this image (\citealt{Sanders21}).}  
    \label{fig:image}
  \end{minipage}
\end{figure*}


Fig.~\ref{fig:image} (upper left) shows an exposure-corrected image of
Abell\,2146 for the energy range $0.5$--$4\keV$.  The subcluster's
cool core is the brightest and densest region in the cluster and is
trailed by a long tail of ram pressure stripped gas that extends over
$200\kpc$.  Based on this extended tail and the bow shock location to
the SE, the subcluster is currently travelling SE.  The primary
cluster lies primarily to the NW of the subcluster's tail with a
second shock to the far NW.  A comparison of the X-ray and optical
images in Fig.~\ref{fig:image} (upper left and right) reveals the
separation of the subcluster (SE) and primary cluster galaxies (NW)
with the bulk of the X-ray emission located between them.  Based on
the X-ray structure, galaxy distribution and hydrodynamical
simulations, Abell\,2146 is a collision between two clusters observed
${\sim}0.1\Gyr$ after the subcluster passed through the primary
cluster's centre.

The merger axis likely runs approximately NW to SE through the centre
of the two galaxy distributions.  The primary cluster's cool core has
been destroyed in the collision and the remains have spread
perpendicular to the merger axis to the SW (referred to in
\citet{Russell12} as the SW plume).  Additional plumes are revealed in
the new observations to the E of the subcluster's core (E plume) and
NW of the primary cluster's main remnant (parallel plumes).  These
structures are clearly seen in the RGB image, which covers a larger
field of view and reveals the more extended cluster atmosphere
including the preshock gas beyond each shock front
(Fig. \ref{fig:image}).

Hydrodynamical simulations show that this was likely an off-axis
merger with the two cluster cores passing ${\sim}100\kpc$ from each
other at closest approach.  This scenario is consistent with the asymmetry in
the primary core remnant about the merger axis
(\citealt{Chadayammuri22}).  Dynamical analyses using
galaxy line of sight velocities indicate that the merger axis is
inclined at only $13$--$19\deg$ to the plane of the sky.  This orientation is
consistent with the clear shock front detections in the X-ray
observations (\citealt{Canning12}; \citealt{White15}).

The subcluster's dense core drives a broad bowshock, ${\sim}500\kpc$
across, seen as a sharp edge in the X-ray surface brightness
${\sim}150\kpc$ ahead (SE) of the leading edge of the core.  The
second shock front, the upstream shock, is located at the far NW edge
of the primary cluster and is propagating in the opposite direction.
The upstream shock forms when material stripped from the subcluster's
core is swept upstream and collides with remaining infalling material.
Hydrodynamical simulations indicate that the structure in this region
is complex and sensitive to the merger parameters.  The upstream shock
evolves rapidly.  Its curvature is variable as it propagates through a
clumpy medium falling in behind the subcluster.  In addition,
simulations show additional shocks and shock-heated structures in this
region (e.g. \citealt{Chadayammuri22}).

\begin{figure}
  \centering
  \includegraphics[width=0.9\columnwidth]{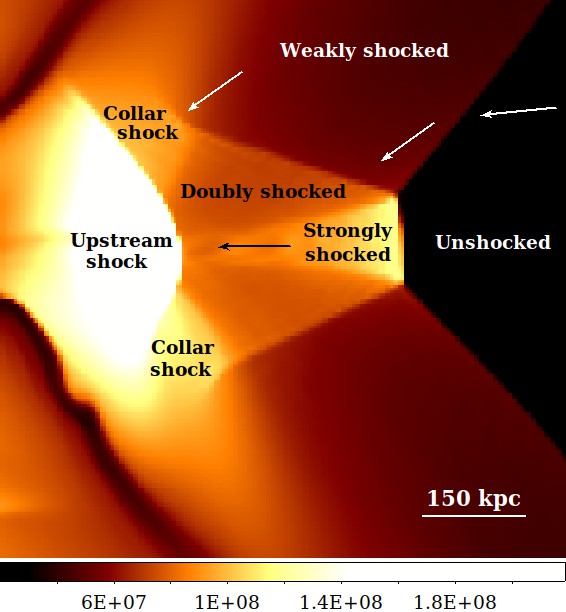}
  \caption{Temperature slice from the hydrodynamical simulations by \citet{Chadayammuri22} showing the upstream shock structure in detail.  The merger axis is oriented left to right and the upstream shock propagates from left to right.  The arrows show the approximate flow direction at various positions.  The colour bar shows the temperature in K.}
  \label{fig:sim}
\end{figure}

Fig. \ref{fig:sim} shows the main features of the complex flow that
feeds the upstream shock in a temperature slice from the hydrodynamic
models of \citet{Chadayammuri22}.  While the slice is taken from their
best matching model, these features are robust to parameter changes,
as can be seen in Fig. 9 of that paper.  The coolest material at the
right hand edge is unshocked gas from the subcluster, which falls
supersonically towards the primary cluster.  This material flows
leftward and converges toward the merger axis under the primary
cluster's gravitational potential.  This results in a series of
strong, normal and weaker, inclined shock fronts.  The upstream shock
structure is expected to be particularly complex.  In addition, it is
projected onto the break up of the primary cluster's core.



Structure in the X-ray surface brightness edges can be highlighted by
subtracting smoothed images.  Fig.~\ref{fig:image} (lower left) shows
an unsharp-masked image, where an exposure-corrected image smoothed
with a 2D Gaussian of $\sigma=10\asec$ has been subtracted from a
similarly smoothed image where $\sigma=2.5\asec$.  The order of
magnitude drop in surface brightness across the contact discontinuity
or cold front on the leading edge of the subcluster core appears
particularly prominent.  The primary core remnant is fully revealed
and extends ${\sim}250\kpc$ from the merger axis.  On the
opposite side of the merger axis, the deeper observation reveals an E
plume that curves from the end of the subcluster's tail almost to the
E edge of the subcluster's core.  Hydrodynamical simulations
suggest that this structure is unrelated to the break up of the
primary cluster's core (\citealt{Chadayammuri22}) but instead may
be a large turbulent eddy.

The deeper observation also shows new structure through the primary
cluster.  The bright nib on the leading edge of the upstream shock is
coincident with the primary cluster's brightest cluster galaxy (BCG-B).
Gas clumps attached to and stripped from the massive galaxies in the
primary cluster lie between the upstream shock and the end of the
subcluster's tail.  To the SW of the collision site, two further
plumes extend perpendicular to the merger axis and parallel to the
primary core remnant.  These plumes are also visible in the raw image.



\section{Spatially resolved spectroscopy}
\label{sec:spec}

\begin{figure*}
  \begin{minipage}{\textwidth}
    \centering
    \includegraphics[width=0.45\columnwidth]{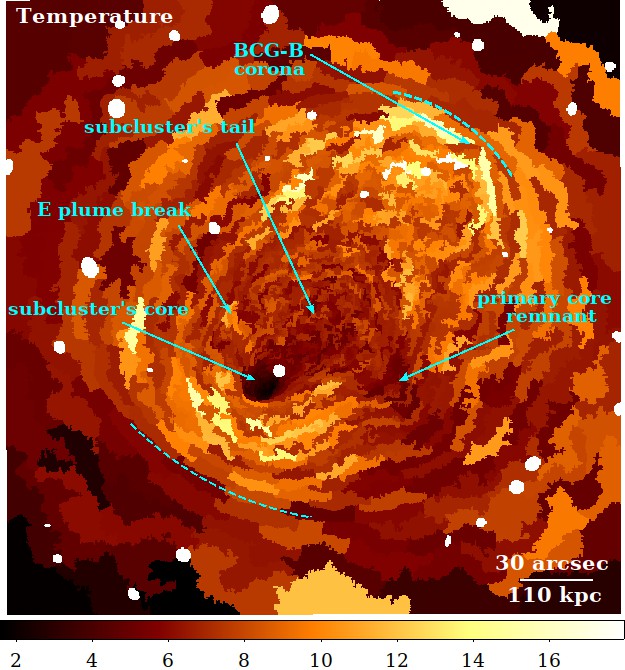}
    \includegraphics[width=0.45\columnwidth]{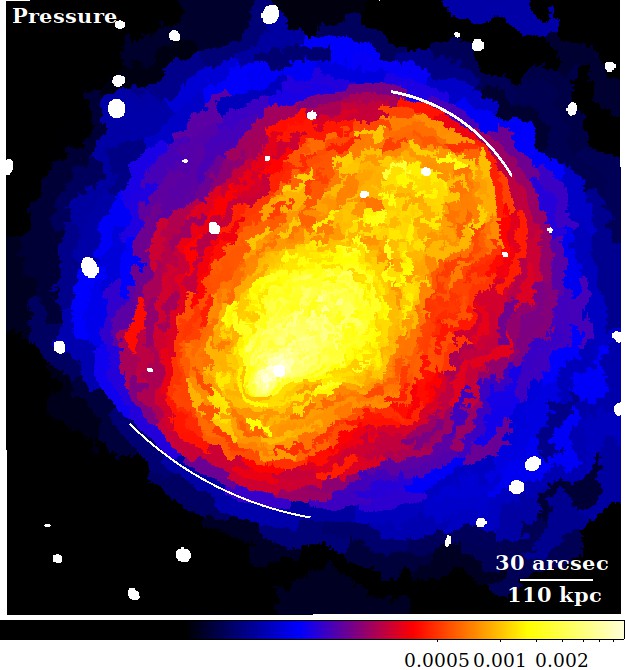}
    \caption{Left: Projected temperature map (keV).  Right: Projected pseudo-pressure map ($\pseudopressure$).  The excluded point sources are visible as small white circles.  The shock front locations determined from surface brightness profile fitting are shown by the arcs.}
    \label{fig:maps}
  \end{minipage}
\end{figure*}

\begin{figure*}
  \begin{minipage}{\textwidth}
    \centering
    \includegraphics[width=0.45\columnwidth]{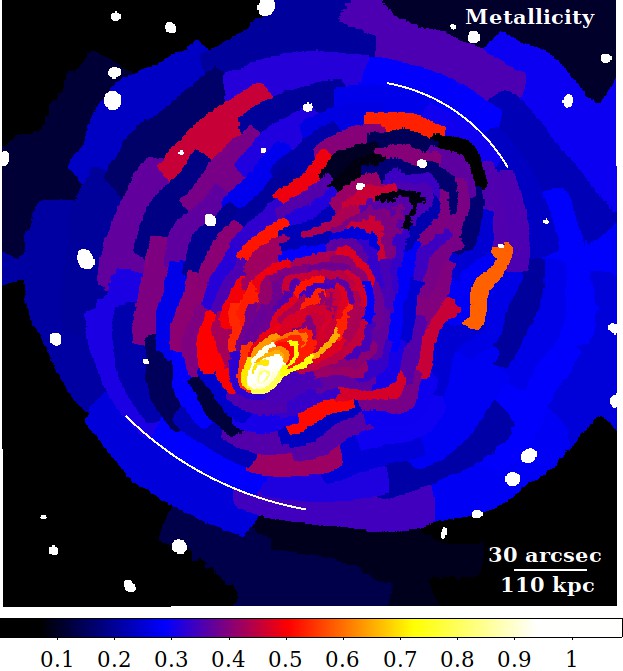}
    \includegraphics[width=0.45\columnwidth]{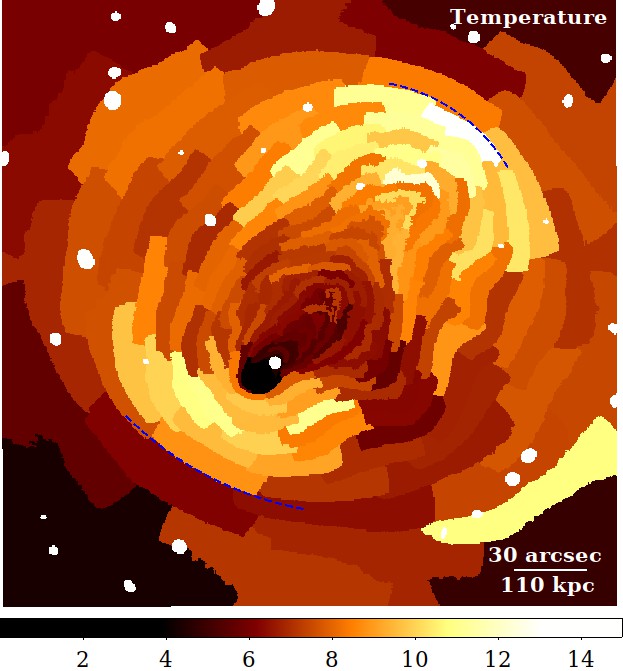}
  \caption{Left: Metallicity map (Z$_{\odot}$).  Right: Projected temperature map (keV) for matching spatial binning.  The excluded point sources are visible as small white circles.}
  \label{fig:metalmap}
  \end{minipage}
\end{figure*}

Detailed maps of the gas temperature, metallicity, normalization and
pressure were produced with a contour binning algorithm
(\citealt{Sanders06}).  Contour binning generates spatial regions by
grouping together neighbouring pixels with similar surface brightness.
Pixels are added to each region until a specified signal-to-noise
ratio is reached.  The dimensions of the regions were restricted so
that their length was at most two and a half times their width.

For each observation, spectra were extracted from each contour binning
region and appropriate responses were
generated.  To ensure sufficient counts, a single blank sky background
for each dataset was extracted from an on-axis region with radius
$1.5\arcmin$.  The spectra were grouped to ensure a minimum of 1 count
per spectral channel.  For each region, spectra from all observations
were fit simultaneously over the energy range $0.5$--$7\keV$ in
\textsc{xspec} v.12.11.1 (\citealt{Arnaud96}) with an absorbed thermal
plasma emission model (\textsc{phabs(apec)}; \citealt{Smith01}).

X-ray spectral modelling of cluster plasma is particularly
straightforward (for a review see e.g. \citealt{Bohringer10}).  All
radiative processes depend on the collision of an electron and an ion.
Due to the low density, all excited ions radiatively de-excite before
another collision occurs.  The low density ensures all photons escape
the cluster.  Radiative transfer calculations are not required.

Thermal plasma emission models, such as \textsc{apec}, are generated
by summing over all electron ion collision rates.  The collision rates
are dependent on the temperature, electron density and ion density.
The normalization of the spectrum is proportional to the electron and
ion densities.  The shape of the spectrum is determined by the
electron temperature and heavy element abundances, typically measured
from line emission.  At temperatures above a few keV, the electron
temperature is predominantly constrained by the exponential dropoff in
the bremsstrahlung continuum shape at high energies.  In contrast, the
emission in a low energy band, such as $0.5$--$4\keV$, is largely
independent of temperature. We note that for gas temperatures below
${\sim}3\keV$, the prominence of Fe L emission at ${\sim}1\keV$
provides a sensitive temperature diagnostic.  The \textsc{apec} model
therefore has three free parameters: temperature, metallicity and
normalization.  Additional parameters for the redshift and absorption
were fixed to 0.234 and the Galactic value of column density
$n_{\mathrm{H}}=3.0\times10^{20}\pcmsq$ (\citealt{Kalberla05}),
respectively.

This model assumes that the electrons are in thermal
Maxwellian equilibrium and the ions are in thermal ionization
equilibrium.  While it is true that on microphysical scales kinetic
effects and temperature anisotropies are important (as attested by lab
experiments and heliospheric observations), on astronomical scales the
fluid description is well-motivated. Kinetic simulations of shocks
(e.g. \citealt{Caprioli14}; \citealt{Park15}) show that
distributions are indeed Maxwellian when integrated on scales of
hundreds of ion gyroradii (${>}10^{12}\cm$ for the ICM).  For
\textit{Chandra's} CCD spectral resolution (${\sim}100\eV$), differences
in the electron and ion temperatures will not significantly affect the
measured parameters.  In future, X-ray microcalorimeters (onboard
e.g. Athena, \citealt{Nandra13}) with spectral resolution of a few eV will detect
thermal line broadening and thereby separately constrain the ion
temperature (\citealt{Hitomi16}).  Similarly, detailed measurements of
the intensity ratios of Fe K$\alpha$ lines with X-ray
microcalorimeters will be able to detect a non-equilibrium ionization
state (e.g. \citealt{Akahori10,Akahori12}; \citealt{Wong11}).

We therefore proceed with the absorbed thermal plasma emission model as
described and determine the best-fit spectral model by minimizing the
C-statistic (\citealt{Cash79}).  The best-fit spectral parameters were
painted onto the contour bin regions to make maps in temperature and
metallicity.  Pseudo-pressure was generated by multiplying
the temperature and square root of the normalization.

For maps with S/N$=32$ (Fig.~\ref{fig:maps}), the uncertainties are
typically ${\sim}15\%$ in temperature, ${\sim}8\%$ in normalization
and ${\sim}18\%$ in pressure.  Regions at lower temperatures, such as
the subcluster's core and tail, have smaller uncertainties in
temperature at $5\%$ and $10\%$, respectively.  This improvement is
due to the temperature sensitivity of the Fe L line emission, which is
prominent at low X-ray temperatures.  The uncertainties in temperature
increase to ${\sim}20\%$ in the postshock gas at $8$--$10\keV$.  These
higher temperatures are difficult to constrain because the exponential
cut off in the bremsstrahlung continuum emission lies beyond Chandra's
energy range.

Fig.~\ref{fig:maps} shows the resulting maps of gas temperature and
pressure.  The pressure peaks in the subcluster's cool core and is
elongated along the merger axis toward the upstream shock.  The
temperature map reveals the ${\sim}2\keV$ gas along the leading edge
of the subcluster's cool core and the steady increase to ${\sim}8\keV$
through the ram pressure stripped tail.  The gas behind each shock
front has been heated to ${\sim}10\keV$.  The primary cluster's core
is visible as a relatively cool plume at ${\sim}6\keV$, which is
${\sim}2\keV$ cooler than the surrounding gas.  Similarly, the E plume
is cooler ($6$--$7\keV$) than the surrounding medium at $8\keV$.  The
temperature map indicates a break in the E plume in a hot region at
$9.5\keV$ covering roughly $6\arcsec\times10\arcsec$.  This region is
coincident with a drop in the X-ray surface brightness.  The parallel
plumes also appear cooler than the ambient by ${\sim}0.5\keV$.  The bright nib on the leading edge of the upstream shock, associated with BCG-B, appears cooler than the surrounding shock-heated gas by at least $5\keV$.  This is likely to be the remains of the massive galaxy's hot atmosphere, or corona, that has survived ram pressure stripping.  The
break up and detailed structure of the cool cores will be discussed
further in separate papers.

\begin{figure*}
  \begin{minipage}{\textwidth}
  \centering
  \raisebox{1cm}{\includegraphics[width=0.4\columnwidth]{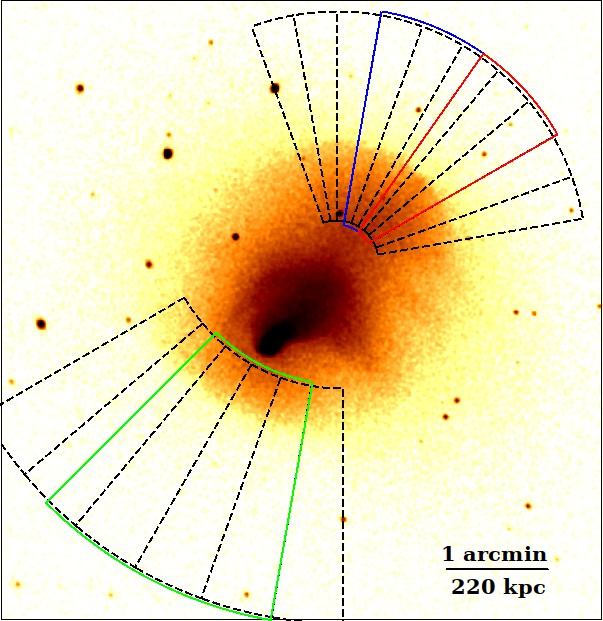}}
  \hspace{1cm}
  \includegraphics[width=0.35\columnwidth]{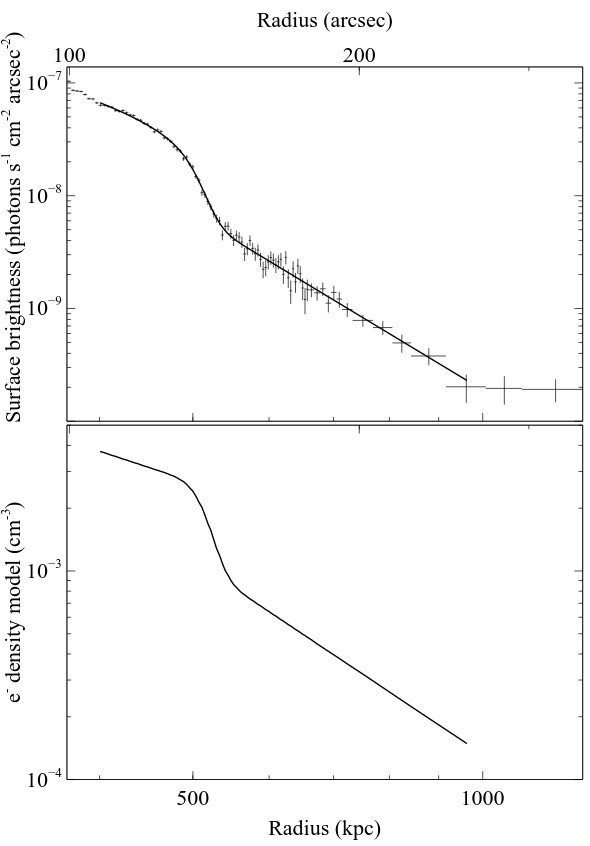}
  \caption{Left: Image with $10^{\circ}$-wide sectors overlaid
      for the bow shock (SE, $210$--$270^{\circ}$) and the upstream shock
      (NW, $10$--$110^{\circ}$), where angles are measured anti-clockwise
      from W.  Note that the overlapping sectors, $+5^{\circ}$ on those shown, are not included for clarity.  Broad sectors for the bow shock (SE, $225$--$260^{\circ}$, green) and the upstream shock (NW, $30$--$55^{\circ}$, red, and $55$--$80^{\circ}$, blue) are also shown with small position offsets for clarity.  Upper right: Surface brightness profile and best-fit model for a sector from $225$--$260^{\circ}$ at the bow shock.  Lower right: Corresponding deprojected electron density model.}
  \label{fig:sectors}
  \end{minipage}
\end{figure*}

For the lower spatial resolution maps with S/N$=72$
(Fig.~\ref{fig:metalmap}), the uncertainties are typically $5\%$ in
temperature and $15$--$20\%$ in metallicity.  Similarly to the S/N$=32$
maps, the uncertainties are lower in the cool core where the Fe L line
emission provides tighter constraints on temperature and metallicity.
The uncertainties are also higher in the postshock regions where the
gas is almost fully ionized and there is little line emission.  This
effect can be clearly seen in Fig.~\ref{fig:metalmap} in the
shock-heated region behind the upstream shock where there are several
regions with very low, essentially unconstrained, metallicity.

Fig.~\ref{fig:metalmap} shows a clear peak at roughly solar
metallicity in the subcluster's cool core.  Metallicity is enhanced at levels of
$0.4$--$0.6\Zsun$ through the ram-pressure stripped tail and the E plume.  The metallicity is ${\sim}0.25\Zsun$ beyond this for the bulk of the cluster atmosphere.  The
subcluster's cool core has likely been enriched by stellar winds and supernovae in
BCG-A.  This material is then stripped from the cool core and mixes
with the lower metallicity ambient medium to produce a steady decline
in measured metallicity through the tail.  The primary cluster's cool
core does not have such a clear metallicity peak.  The metallicity structure of the
cool cores will be examined in more detail in a separate paper.


\section{Shock structure}
\label{sec:shockstructure}

\begin{figure*}
  \begin{minipage}{\textwidth}
    \centering
    \includegraphics[width=0.3\columnwidth]{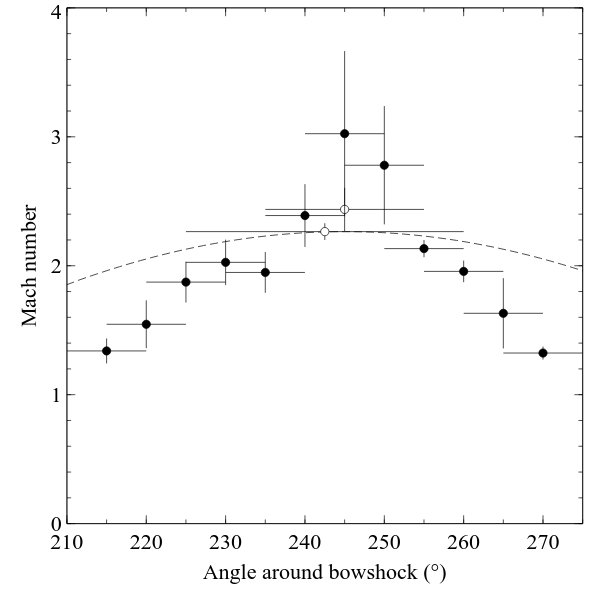}
    \includegraphics[width=0.3\columnwidth]{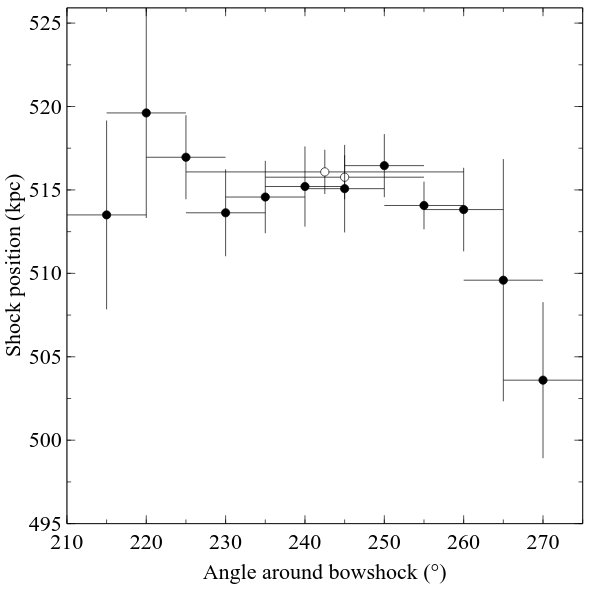}
    \includegraphics[width=0.3\columnwidth]{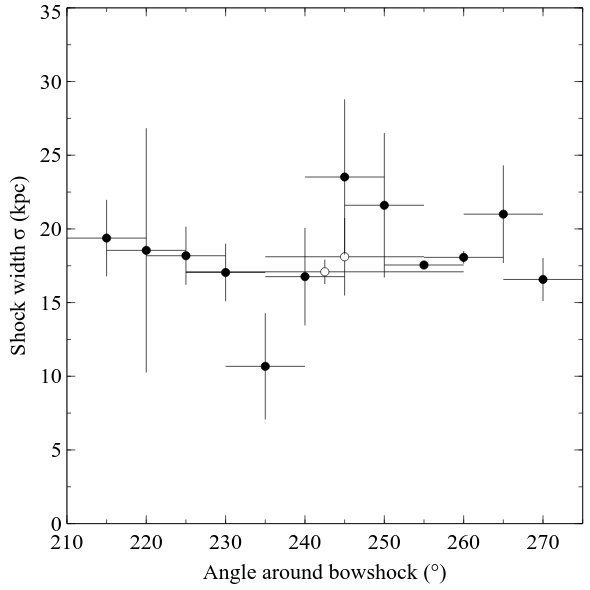}
    \caption{The best-fit
      Mach number, shock position and shock width parameters
      determined from fitting the projected density jump model to
      surface brightness profiles extracted in $10^{\circ}$ sectors
      around the bow shock front (solid points).  The expected $\mathrm{cos}\theta$ dependence of the Mach number (see text) is shown by the dashed line.  The best-fit
      parameters do not vary significantly for a $225$--$260^{\circ}$
      sector or a narrower $235$--$255^{\circ}$ sector (open
      points), which was selected for subsequent analysis of the
      temperature and density structure of the shock.}
    \label{fig:edgepars}
  \end{minipage}
\end{figure*}


\begin{figure}
\centering
\includegraphics[width=0.98\columnwidth]{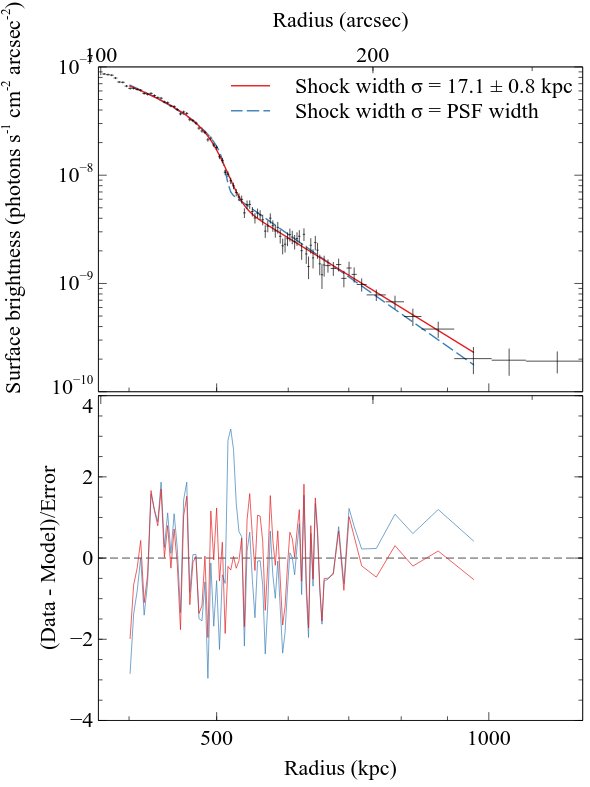}
\caption{Upper: Surface brightness profile across the bow shock front with the best-fit model for a projected density jump of width $\sigma$ (solid red line).  The best-fit model for a projected jump with zero width convolved with a Gaussian with $\sigma=1\asec$ to approximate the \textit{Chandra} PSF is shown for comparison.  Lower: Residuals calculated by subtracting each best-fit model from the observed surface brightness profile and dividing the result by the uncertainties.}
\label{fig:bowSB}
\end{figure}

Fig.~\ref{fig:sectors} shows $10^{\circ}$-wide sectors used to extract
surface brightness profiles across the bow and upstream shock fronts.
Overlapping $10^{\circ}$-wide sectors, offset in angle by $5^{\circ}$
from those shown, were also used.  The regions were positioned to
align with the centre and curvature of each respective surface
brightness edge.  The analysis was repeated with modest offsets in the
region position and the surface brightness profile modelling results
were found to be robust.  Each sector was divided into radial bins of
$1\asec$ width, increasing to $2.5\asec$ and larger for radii greater
than $100\kpc$ beyond the surface brightness edge where the background
dominates.  This procedure ensured at least 30 source counts are
contained in each radial bin for the resulting surface brightness
profiles.  Point sources were excluded.  The background was subtracted
using the surface brightness measured in source-free regions beyond
each sector.  The energy range was restricted to $0.5$--$4\keV$ to
minimise the temperature dependence of the profile (section
\ref{sec:spec}; see e.g. \citealt{Churazov16}) and maximise S/N at the
shock fronts.  The density jump across each shock front is therefore
determined essentially independent of the temperature change (see
e.g. \citealt{Markevitch07}).

Each surface brightness profile was fitted with a model for a
projected density jump (Fig.~\ref{fig:sectors}).  This model consists
of a power law for the preshock gas density, a second power law for
the postshock gas density and a sharp density jump at the shock edge
convolved with a Gaussian function of width $\sigma_{\mathrm{sh}}$.
The model has six free parameters: the slopes ($\alpha_1$, $\alpha_2$)
and normalizations ($\rho_1$,$\rho_2$) of each power law in electron
density, the position of the shock, $R_{\mathrm{sh}}$, and the width
of the shock, $\sigma_{\mathrm{sh}}$.  The subscripts 1 and 2 denote
the preshock (upstream) and postshock (downstream) values,
respectively.  The density model was projected along the line of sight
by assuming the same shock curvature along the line of sight as
measured in the plane of the sky.  This model was then fitted to the
observed surface brightness profiles by minimizing $\chi^2$.

Fig.~\ref{fig:sectors} shows the density model and the corresponding
surface brightness model overlaid on a surface brightness profile
extracted across the bow shock for the broad sector from $225$ to
$260^{\circ}$.  The projected density jump model is an excellent fit
in this sector with $\chi^{2}=75$ for 77 degrees of freedom.  The
uncertainties on each parameter are calculated by evaluating the
best-fit model and parameters for 1000 Monte Carlo realizations of the
surface brightness profile.  Increasing or decreasing the subtracted background
level by $1\sigma$ does not significantly alter the measured
parameters.  From the best-fit density jump, the shock Mach number is
then

\begin{equation}
  M=\left(\frac{2\left(\rho_2/\rho_1\right)}{\gamma + 1 - \left(\rho_2/\rho_1\right)\left(\gamma-1\right)}\right)^{1/2},
  \label{eq:M}
\end{equation}

\noindent where the adiabatic index $\gamma=5/3$ for a monatomic gas.


The density model assumes a spherical and steady shock
front propagating in the plane of the sky and power-law profiles in
gas density in the pre- and postshock gas.  This model is most
applicable close to the shock front.  Therefore, we set an inner radial
limit of $400\kpc$ which excludes substructure around the
subcluster's core.  Based on galaxy dynamical measurements and
hydrodynamical simulations, the merger axis in Abell\,2146 is estimated
to be only $13$--$19^{\circ}$ from the plane of the sky
(\citealt{Canning12}; \citealt{White15}; \citealt{Chadayammuri22}).
This angle is entirely consistent with the clear detection of two sharp
shock fronts, which would appear smeared in projection at larger
inclination angles.  Inclination effects are therefore expected
to be minimal.

\subsection{Bow shock front}
\label{sec:bowwidth}

\begin{figure*}
  \begin{minipage}{\textwidth}
    \centering
    \includegraphics[width=0.45\columnwidth]{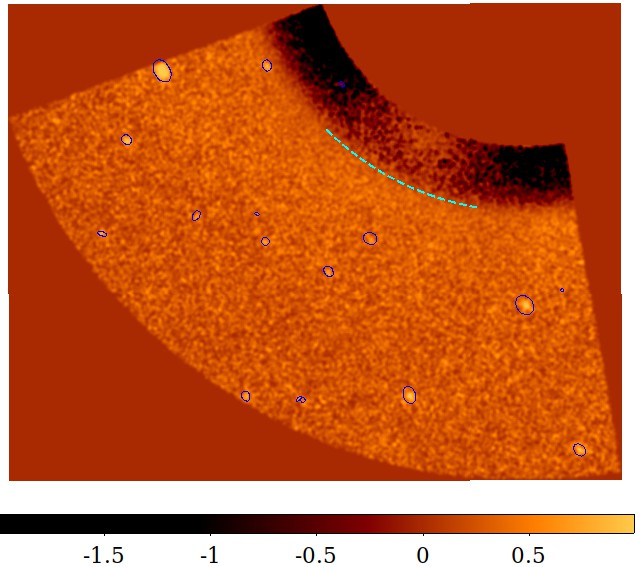}
    \includegraphics[width=0.45\columnwidth]{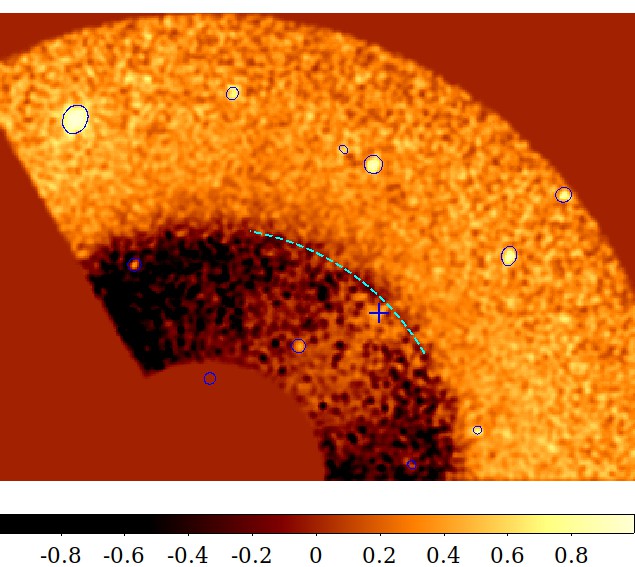}
    \caption{Exposure-corrected images in the $0.5$--$4\keV$ energy band for the shock sectors with the best-fit projected density jump model subtracted.  The residual images were then normalized by the exposure-corrected image.  The colour bar values therefore represent the fractional difference (data - model)/data.  Left: Bow shock region where the model was determined for the $225$--$260^{\circ}$ sector.  Right: Upstream shock region where the model was determined for the $55$--$80^{\circ}$ sector.  Point sources are marked by blue circles and the shock locations are marked by cyan dashed lines.  The location of BCG-B is shown by the blue cross (Fig.~\ref{fig:image}).}
    \label{fig:subshockmodel}
  \end{minipage}
\end{figure*}

Fig.~\ref{fig:edgepars} shows the best-fit Mach number, shock radius
and shock width as a function of angle around the bow shock front.  As
expected for an archetypal bow shock, the Mach number peaks at the
centre or `nose' of the shock front ($235$--$255^{\circ}$) at
$2.44\pm0.17$.  The Mach number declines with angle to $M{<}2$, where
the shock becomes oblique.  The decline in Mach number appears
symmetric; values at equal distances from the shock `nose' are
consistent within the uncertainties.  If the shock is steady, and the
velocity of the preshock medium is uniform, the shock Mach number is
expected to vary as $\mathrm{cos}\theta$, where $\theta$ is the angle
between the velocity of the subcluster's core and the normal to the
shock front (in Fig.~\ref{fig:edgepars}, $\theta=0$ at approximately
$245\deg$).  Fig.~\ref{fig:edgepars} shows that the angular dependence
is much steeper than this expectation.  This indicates that the
shock velocity is not steady and the preshock medium is not uniform.  We note
that if the motion of the subcluster's core was not in the plane of
the sky, the angular dependence would be shallower so this does not
explain the discrepancy.

In general, the measured shock properties, including the Mach number,
do not vary significantly for a broad sector across the centre of the
shock front ($225$--$260^{\circ}$, Fig.~\ref{fig:sectors} and
\ref{fig:edgepars}).  This sector is therefore used to evaluate the
imaging and spectral parameters across the shock front because it
probes the narrow region of the normal shock and captures a sufficient
number of photons.  We note that an even more conservative sector selection of
$235$--$255\deg$ produces consistent results.

For the first time, we resolve and measure the width of a cluster
merger shock front.  For the $225$--$260^{\circ}$ sector, the width is
$17\pm1\kpc$ and measurements in all sectors are consistent with
this value within the uncertainties.  Fig.~\ref{fig:bowSB} shows that this width is clearly preferred over an unresolved width with $\chi^2=75$ for 77 degrees of freedom compared to $\chi^2=100$ for 78 degrees of freedom.  Similarly, the position of the
shock front is consistent in all sectors, $R_{\mathrm{s}}=516.3\pm1.3\kpc$, although the uncertainties
increase significantly at large angles where the surface brightness
declines.  Fig.~\ref{fig:bowSB} shows that the Mach number in this
large sector $M=2.24\pm0.09$ from the density jump $r=2.50\pm0.08$ (Eq. \ref{eq:M}).  The best-fit shock width of
$\sigma_{\mathrm{sh}}=17\pm1\kpc$ is significantly preferred over an unresolved width ($\chi^2=75$ for 77 degrees of freedom compared to $\chi^2=99$ for 78 degrees of freedom).  An unresolved shock front is clearly a poor fit around the model's inflection point with residuals of ${>}3\sigma$ significance at a radius of $520\kpc$ in the lower panel of Fig.~\ref{fig:bowSB}.  

Fig.~\ref{fig:subshockmodel} shows residual images where the best-fit
projected density model for the $225$--$260\deg$ sector has been
subtracted from an image covering the shock sector.  Whilst this
model clearly oversubtracts the postshock emission at large angles
where the density jump is smaller, no other significant
residuals in pre- or postshock gas are seen.  The $225$--$260\deg$ sector appears clear of
substructure.  We proceeded with this region for the spectral
analysis.

\begin{figure}
  \centering
  \includegraphics[width=0.98\columnwidth]{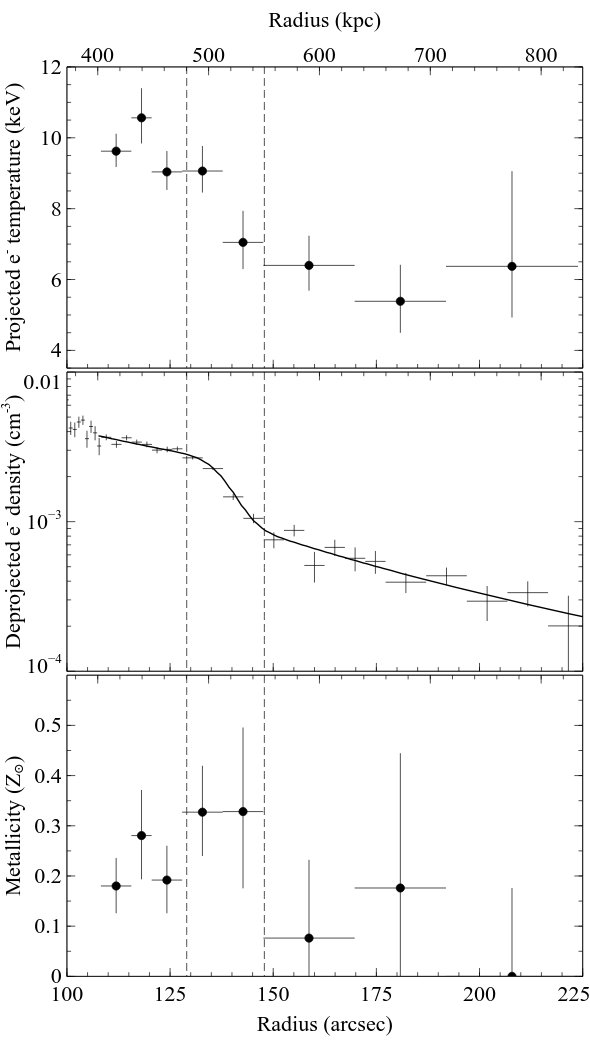}
  \caption{Projected temperature and metallicity profiles extracted in the $225$--$260\deg$ sector across the bow shock front (top and bottom panels).  Deprojected electron density profile with the best-fit density jump model (solid black line) overlaid (middle panel).  Vertical dashed lines approximate the full width of the density jump and guide the radial binning selected for the temperature profile.}
  \label{fig:bowtempprofile}
\end{figure}


Temperature and metallicity profiles were extracted using the bow
shock sector ($225$--$260\deg$; Fig.~\ref{fig:sectors}) and
wider regions each with ${\sim}3000$ counts spread over 75 datasets.  This ensured temperature precision of $10$--$20\%$.  As
detailed in section \ref{sec:spec}, spectra and responses were
extracted from each region and fitted simultaneously with an absorbed
\textsc{apec} model in \textsc{xspec}.  Results were consistent when using different radial binning and different
subsets of obs. IDs, including comparing the 2010
observations and the new data.  Deprojected electron density profiles
were produced using the \textsc{dsdeproj} deprojection routine, assuming
 spherical symmetry to subtract the projected contribution
from each successive annulus (\citealt{SandersFabian07};
\citealt{Russell08}).

Fig.~\ref{fig:bowtempprofile} shows the projected temperature,
metallicity and deprojected density profiles for the bow shock sector.
The shock front is clearly visible as a rapid increase in temperature
from ${\sim}6.0\pm0.6\keV$ to $9.3\pm0.2\keV$.  From the density jump and the sound speed in the preshock gas $c_{\mathrm{s}}=1250^{+80}_{-40}\kmps$, the shock velocity $v_{\mathrm{sh}}=2800^{+200}_{-100}\kmps$.  A temperature
profile extracted from a sector covering a reduced angular range of
$235$--$255\deg$ produced a consistent result within the uncertainties.
The deprojected electron density also shows the shock front clearly
with a steady increase over ${\sim}40\kpc$.  The increase is consistent with the
best-fit width with $\sigma_{\mathrm{sh}}=17\kpc$.  The best-fit metallicity is
roughly constant through this sector at $0.24\pm0.03\Zsun$ in the
postshock gas and $0.1\pm0.1\Zsun$ in the preshock gas.  This difference is not significant given the uncertainty on the preshock metallicity.  The
temperature profile is analysed in more detail in section
\ref{sec:equilibration} where we measure the electron-ion thermal
equilibration timescale.

\begin{figure}
  \centering
  \includegraphics[width=0.98\columnwidth]{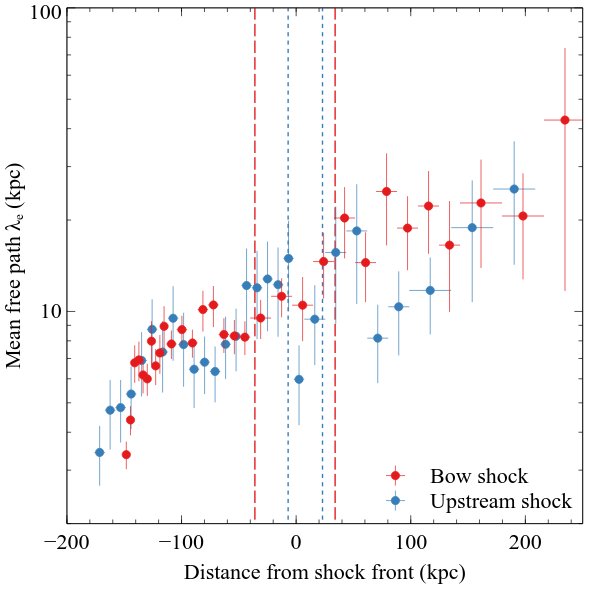}
  \caption{Electron mean free path across the bow and upstream shock fronts.}
  \label{fig:meanfreepath}
\end{figure}

The shock width can be compared with the electron mean free path (\citealt{Spitzer56})

\begin{equation}
  \lambda_{\mathrm{e}}{\sim}11\kpc\left(\frac{T_{\mathrm{e}}}{9\keV}\right)^{2}\left(\frac{n_{\mathrm{e}}}{2.3\times10^{-3}\pcmcu}\right)^{-1},
\end{equation}

\noindent where $T_{\mathrm{e}}$ is the electron temperature and
$n_{\mathrm{e}}$ is the electron density.  For the bow shock front,
$\lambda_{\mathrm{e}}=11\pm2\kpc$ (Fig.~\ref{fig:meanfreepath}), which
is similar to the shock width.  A collisional shock should have a
width roughly a few times the mean free path.  Any deformation of the shock
shape across the large sector analysed would increase the
width further.  The bow shock therefore appears too narrow for a collisional shock.

Instead, the bow shock is likely a collisionless shock with width of
order the ion gyroradius (typically npc).  Small scale turbulent
eddies in the preshock region will warp the shape of this narrow
shock.  These local gas motions modulate the shock speed to produce an
uneven shock surface, which appears broader when seen in projection.

The width of the shock front will grow due to turbulence if the shock
front is not driven by the subcluster's core and instead propagates
freely (\citealt{Nulsen13}).  During the infalling leg of the merger,
the core accelerates and drives the shock front.  The standoff
distance of the shock agrees with the expectation for a steady shock
(\citealt{Zhang19}).  For a $M{\sim}2$ shock, that distance is
${\sim}0.4$ times the radius of the core.  Pressure fluctuations
propagate slowly into the high density subcluster's core compared to
the surrounding hot atmosphere.  The leading edge of the subcluster's
core is essentially rigid.  During infall, velocity perturbations on
scales comparable to or greater than the subcluster's core radius will
cause only modest fluctuations in the shock standoff distance.
Perturbations on smaller scales can cause local displacements in the
shock front but these are limited in magnitude by the pressure
gradient behind the shock.  Therefore, before core passage, perturbations in the
shock's shape due to turbulence are modest.

After the shock detaches from the leading edge of the subcluster's
core, it is no longer driven by a rigid body and displacements in it
can accumulate.  The observed standoff distance for the bow shock is
several times larger than the radius of the core.  The merger is
therefore observed after core passage.  The shock has detached and
propagates almost freely.

The rms radial displacements in the shock front
${\Delta}R=\sqrt{Rl}\left(\sigma_{\mathrm{turb}}/v_{\mathrm{sh}}\right)$,
where the turbulence rms speed is $\sigma_{\mathrm{turb}}$, the
coherence length $l$ is typically ${\sim}0.1R$ and $v_{\mathrm{sh}}$
is the shock speed (\citealt{Nulsen13}).  For a shock radius
$R{\sim}500\kpc$, width $\sigma{\sim}20\kpc$ and shock velocity
$v_{\mathrm{sh}}=2800^{+200}_{-100}\kmps$, we estimate
$\sigma_{\mathrm{turb}}=290\pm30\kmps$.  This value measured at a
radius of a few hundred kpc is consistent with turbulence measured in
cluster cores to radii of a few tens of kpc (e.g. \citealt{Sanders11};
\citealt{Hitomi16}).  This implies a slow increase in turbulent
velocity with radius.

\subsection{Upstream shock front}

\begin{figure*}
  \begin{minipage}{\textwidth}
    \centering
    \includegraphics[width=0.3\columnwidth]{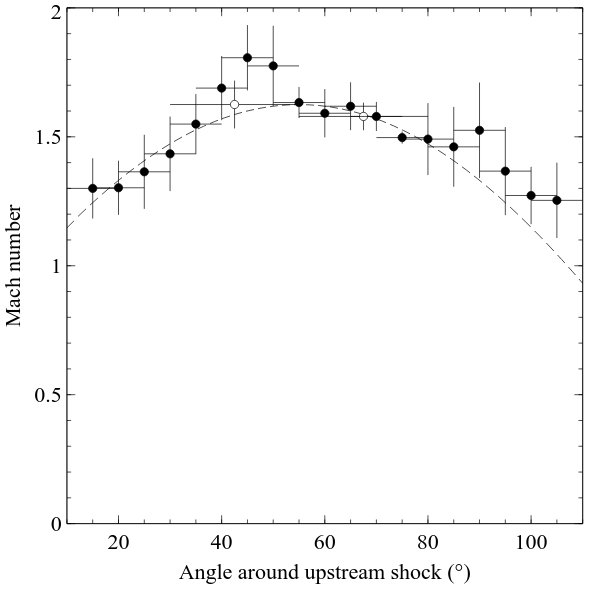}
    \includegraphics[width=0.3\columnwidth]{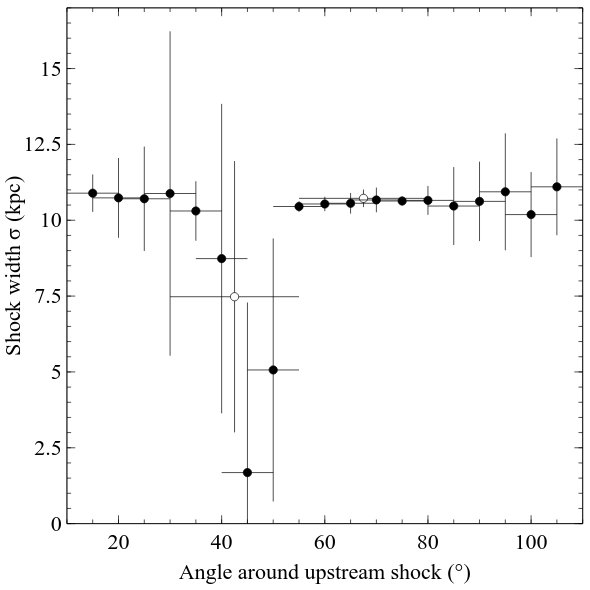}
    \includegraphics[width=0.32\columnwidth]{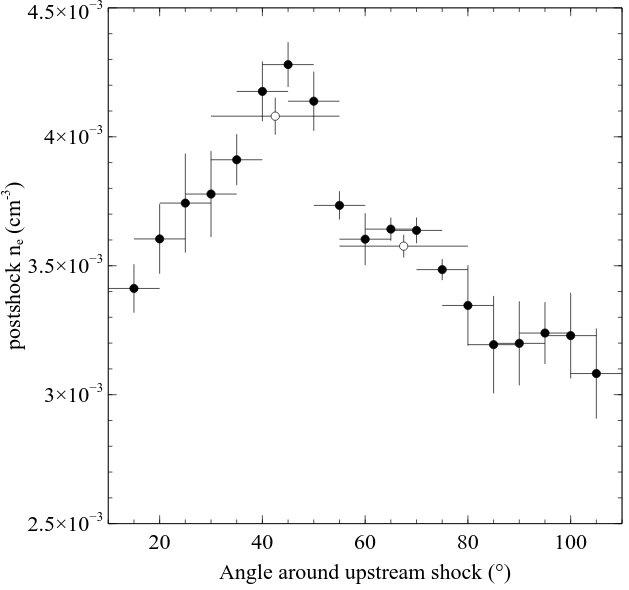}
    \caption{The best-fit Mach number, shock width and postshock electron density parameters determined from fitting the projected density jump model to surface brightness profiles extracted in $10\deg$ sectors around the upstream shock front.  The expected $\mathrm{cos}\theta$ dependence of the Mach number (see text) is shown by the dashed line.  The $30$--$55\deg$ sector contains trails of ram pressure stripped gas from the massive galaxies in the primary cluster.}
    \label{fig:upstreamedgepars}
  \end{minipage}
\end{figure*}

\begin{figure*}
\begin{minipage}{\textwidth}
\centering
\includegraphics[width=0.7\columnwidth]{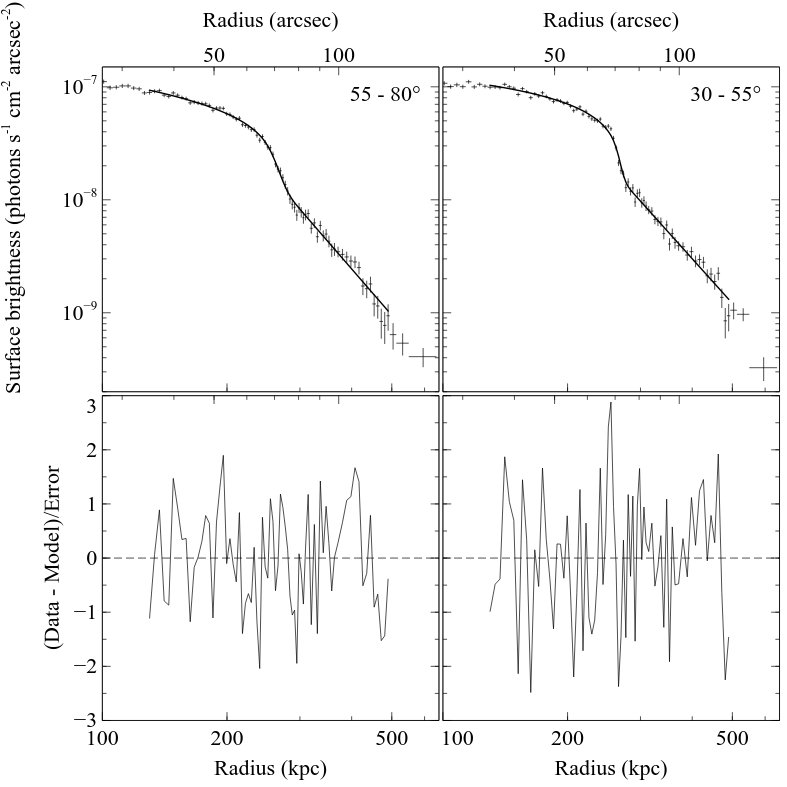}
\caption{Upper: Surface brightness profiles across the upstream shock front for two sectors ($30$--$55\deg$ and $55$--$80\deg$) with the best-fit model for a projected density jump (solid lines).  Lower: Residuals calculated by subtracting the best-fit model from the observed surface brightness profile and dividing the result by the uncertainties.}
\label{fig:upstreamSB}
\end{minipage}
\end{figure*}

Using the $10\deg$-wide sectors shown in Fig.~\ref{fig:sectors}, we
extracted a series of surface brightness profiles across the upstream
shock front and fit each with the projected density model described in
section \ref{sec:shockstructure}.  Fig.~\ref{fig:upstreamedgepars}
shows the best-fit Mach number, shock width and postshock density as a
function of angle around the upstream shock front.  Similar to the bow
shock front, the Mach number decreases with angle around the shock
from a peak at $M=1.58\pm0.05$.  The decline in Mach number with angle
is shallower compared to the bow shock and appears to match the
expected $\mathrm{cos}\theta$ dependence much more closely
(Fig.~\ref{fig:upstreamedgepars}).  For the upstream shock, the Mach
number drops by $15\%$ over $30^{\circ}$ compared to ${\sim}45\%$ for
a similar angular range across the bow shock.  Hydrodynamical
simulations show the structure of the upstream shock front is complex,
rapidly evolving, and sensitive to the merger scenario
(\citealt{Chadayammuri22}; section \ref{fig:image}).  Therefore, it
seems unlikely that agreement with theoretical expectations is a
result of a steady flow and uniform velocity in the preshock medium.


Emission from the galaxy atmospheres, particularly BCG-B, produces a
clear artificial spike in the best-fit Mach number and postshock
electron density from $30$ to $55\deg$.  Comparison of the surface
brightness profiles from the $30$--$55\deg$ and $55$--$80\deg$ sectors
(Fig.~\ref{fig:upstreamSB}), and examination of the residual image
(Fig.~\ref{fig:subshockmodel}), indicate that ram pressure stripped
material from the BCG's halo is spread over at least $100\kpc$
throughout this sector.  Additional contributions from the halos of
other large galaxies in the primary cluster are also likely
(Fig.~\ref{fig:image}).  It was therefore not possible to cleanly mask
out this emission in the $30$--$55\deg$ sector.  The projected density
model was a poor fit in this sector, $\chi^{2}=123$ for 73 degrees of
freedom compared to $\chi^{2}=68$ for 73 degrees of freedom in the
$55$--$80\deg$ sector (Fig.~\ref{fig:upstreamSB}).  The galaxy halo of
BCG-B is also detected as a cool patch in the middle of the upstream
shock front in the temperature map (Fig.~\ref{fig:maps}).  We
therefore proceed by analysing the upstream shock parameters primarily
in the $55$--$80\deg$ sector.  An analysis of the break up of the cool
cores and galaxy halos in the merger will be published separately.

Using the $55$--$80\deg$ sector (Fig.~\ref{fig:sectors}), we
resolve the width of the upstream shock front.  With best-fit width
$\sigma_{\mathrm{sh}}=10.7\pm0.3\kpc$, the upstream shock is significantly narrower
than the bow shock.  The upstream shock is sharper than the
bow shock even in the raw images (Fig.~\ref{fig:image}), which is
consistent with this result.  The measured shock width is consistent
in all $10^{\circ}$ sectors, although the uncertainty is particularly
large from $30$--$55^{\circ}$ as expected given the substructure (Fig.~\ref{fig:upstreamedgepars}). The Mach number in this large sector $M=1.58\pm0.05$ from the best-fit density jump $r=1.82\pm0.07$ (Eq.~\ref{eq:M}).


\begin{figure}
\centering
\includegraphics[width=0.98\columnwidth]{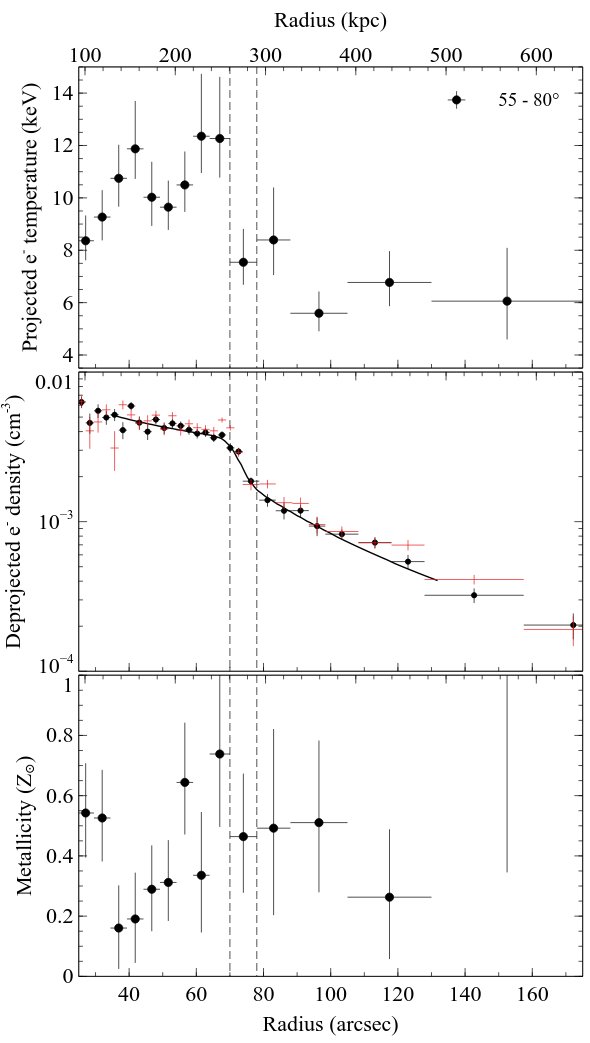}
\caption{Projected temperature and metallicity profiles across the upstream shock front in the $55$--$80\deg$ sector (top and bottom panels).  Deprojected electron density profile (middle panel) with the best-fit density jump model (solid black line) and the result from the $30$--$55\deg$ sector overlaid (red points).  Vertical dashed lines approximate the full width of the density jump and guide the radial binning selected for the temperature profile.}
  \label{fig:upstreamtempprofile}
\end{figure}

Fig.~\ref{fig:upstreamtempprofile} shows the projected temperature,
metallicity and deprojected electron density profiles across the
upstream shock front in the $55$--$80\deg$ sector.  The temperature
increases at the shock front from $6.0\pm0.6\keV$ at large radius, to
$8^{+2}_{-1}\keV$ roughly $50\kpc$ ahead of the shock front to
$12.3^{+1.7}_{-1.0}\keV$ immediately behind the shock front.  Although the
increase ahead of the shock front appears only modestly significant,
we show in section \ref{sec:equilibration} that this increase in the
preshock gas temperature explains the anomalously high postshock
temperature, noted in our previous work
(\citealt{Russell12}).  It is therefore likely that a ${\sim}2\keV$ increase in temperature occurs ${\sim}50\kpc$ ahead of the upstream
shock front.  From the density jump and the sound speed in the preshock gas $c_{\mathrm{s}}=1450^{+180}_{-90}\kmps$, the shock velocity $v_{\mathrm{sh}}=2300^{+300}_{-200}\kmps$.  No significant increase in the
density is seen at this location, and substructure in the surface brightness profiles has low significance (Fig.~\ref{fig:upstreamSB}).  Although the $30$--$55\deg$ sector shows a
clear increase in the gas density in this region at ${\sim}300\kpc$,
this is coincident with another primary cluster galaxy halo and
therefore unrelated.  Similar to the bow shock front, the metallicity
is approximately constant at $0.39\pm0.15\Zsun$ in the preshock gas
and $0.55\pm0.12\Zsun$ in the immediate postshock gas.

As noted earlier, hydrodynamical simulations of Abell\,2146 have shown
complex structure in the upstream shock (\citealt{Chadayammuri22}; section \ref{sec:image}).  It is likely
that a series of additional weaker shocks lie ahead of the
upstream shock front.  These structures could explain the observed
temperature increase.  Although the temperature increase is not
clearly associated with a sharp density jump, surface brightness
profiles for sectors across the centre of this shock show structure at
low significance.  This region is also projected onto the break up of
the primary cluster's core, which may obscure the shock density
structure.  If the temperature increase was due to a shock precursor,
we would expect a comparable feature ahead of the bow shock front,
which is not observed.  We therefore consider a precursor origin to be less likely.

Following our analysis of the bow shock width above (section
\ref{sec:bowwidth}), we consider the width of the upstream shock
front.  For an upstream shock radius $R{\sim}267\pm1\kpc$, width $\sigma_{\mathrm{sh}}=10.7\pm0.3\kpc$
and shock velocity $v_{\mathrm{sh}}=2300^{+300}_{-200}\kmps$, we
estimate $\sigma_{\mathrm{turb}}=290\pm30\kmps$.  The turbulent level is therefore similar ahead of both shock fronts.  In this model, the
upstream shock is narrower than the bow shock because it is younger.
The upstream shock forms later during the merger, which is confirmed
by hydrodynamical simulations (\citealt{Chadayammuri22}).

The electron mean free path across the upstream shock is very similar
to the bow shock at $\lambda_{\mathrm{e}}=15\pm5\kpc$
(Fig.~\ref{fig:meanfreepath}).  The narrower width of the upstream
shock relative to the mean free path strengthens our earlier conclusion that the shock fronts in
Abell\,2146 are collisionless and broadened in projection by turbulence in the
preshock gas.

\vspace{\baselineskip}

In summary: we fit a projected density jump model to surface
brightness profiles extracted in narrow sectors across each shock.
The model's free parameters include the shock width and the density increase across the shock,
which is used to calculate the Mach number.  The bow shock Mach number
$M=2.24\pm0.09$.  The angular dependence of the Mach number is much
steeper than the expected $\mathrm{cos}{\theta}$ dependence.  This is
likely due to a break down in key assumptions of steady shock velocity
and uniform preshock medium.  The best-fit shock width of
$\sigma_{\mathrm{sh}}=17\pm1\kpc$ is significantly preferred over
an unresolved width.  A similar analysis of the upstream shock front
finds $M=1.58\pm0.05$ and $\sigma_{\mathrm{sh}}=10.7\pm0.3\kpc$.  The
shocks appear too narrow to be collisional.  Collisionless shocks are
typically npc in width but will appear much broader in projection if
their smooth shape is warped by turbulence in the preshock gas.  We
show that both shock widths are consistent with collisionless shocks
propagating through a turbulent medium with
$\sigma_{\mathrm{turb}}=290\pm30\kmps$.  The upstream shock is
narrower than the bow shock because it is younger.  We also find a
${\sim}2\keV$ increase in temperature ${\sim}50\kpc$ ahead of the
upstream shock front, which is likely due to additional weak shocks.


\section{Electron-ion thermal equilibration behind shock fronts}
\label{sec:equilibration}

Electron and ion heating by collisionless shocks remains an unsolved
problem in shock physics.  This problem has received much less
attention than non-thermal particle acceleration
(e.g. \citealt{Ghavamian13}).  The more massive ions are directly
heated in the shock layer by plasma-wave interactions between the ions
and the magnetic field.  The electrons then subsequently thermally
equilibrate with the shock-heated ions.  However, the exact mechanism
and timescale are unknown.  Collisionless shocks occur over a wide
range of scales in astrophysics from accretion shocks at the
intersection of massive cosmic structure filaments to supernova
remnants and solar wind shocks (e.g. \citealt{Marcowith16}).  Whilst
the shock parameters differ, only in galaxy clusters are we able to
map the large-scale equilibration processes in a single observation
(see section \ref{sec:data}) and avoid systematic errors from
cross-calibrating observations in different wavebands.

\subsection{Equilibration models}

Following \citet{Markevitch06} and \citet{Russell12} (see also
\citealt{Wang18}), we measured the temperature profiles across each shock
front to estimate the timescale for the electrons and ions to return to thermal equilibrium.  Fig.~\ref{fig:boweiplot} shows the measured electron temperature
across the bow shock in Abell\,2146 for the sector $225$--$260\deg$.  We
note that an even more conservative sector selection from $235$--$255^{\circ}$
produces a consistent result.  We compare the observed profile
with two models for electron-ion thermal equilibration.  The instant
equilibration model assumes that the electrons rapidly equilibrate
with the hotter ions over an unresolved distance.  The electron
temperature and the ion temperature jump at the shock front to the
post-shock gas temperature given by the Rankine-Hugoniot shock jump
conditions.  The second model is the collisional equilibration model, which assumes an unresolved
adiabatic increase in electron temperature,

\begin{equation}
  T_{\mathrm{e,2}}=T_{\mathrm{e,1}}\left(\frac{n_{\mathrm{e,2}}}{n_{\mathrm{e,1}}}\right)^{\gamma-1}.
\end{equation}

\noindent Charge neutrality ties the electron density to the ion density very rigidly so the electron density jumps at the ion shock.  This adiabatic increase is followed by a slower equilibration via Coulomb
collisions at a rate

\begin{equation}
  \frac{\mathrm{d}T_{\mathrm{e}}}{\mathrm{d}t}=\frac{T_{\mathrm{i}}-T_{\mathrm{e}}}{t_{\mathrm{cc}}}.
  \label{eq:int}
\end{equation}
  
\noindent $T_{\mathrm{e}}$ is the electron temperature, $T_{\mathrm{i}}$ is the ion temperature and $n_{\mathrm{e}}$ is the electron temperature.  The subscripts 1 and 2 denote the preshock and postshock values, respectively.  The Coulomb collisional timescale is given by

\begin{equation}
t_{\mathrm{cc}}=2.54\times10^{8}\left(\frac{T_{\mathrm{e}}}{10^{8}\K}\right)^{3/2}\left(\frac{n_{\mathrm{e}}}{10^{-3}\pcmcu}\right)^{-1}\yr
\end{equation}


\noindent (e.g. \citealt{Wong09}; \citealt{Sarazin16}).  The electron
temperature as a function of time for the collisional model was
determined analytically by integrating equation \ref{eq:int}.  The
shock velocity in the postshock gas was used to determine the electron
temperature as a function of distance.

To summarise: the observed preshock electron temperature was combined
with the measured electron density jump to predict the postshock
electron temperature under the assumptions of two different
equilibration models.  The collisional equilibration model assumes
adiabatic compression of the electrons at the shock followed by
equilibration on the collisional timescale.  The instant equilibration
model assumes an additional electron heating mechanism that rapidly
heats the electrons over an unresolved distance.  For details see
\citet{Russell12}.

Both equilibration models were projected by calculating the emission
measure of gas at each temperature along lines of sight through the
cluster.  The projection assumed the density model determined in
section \ref{sec:shockstructure} and blurring to
  incorporate the shock width.  By simulating spectra in \textsc{xspec}
with the predicted emission measure distribution, we generated
projected temperature profiles for each model.  The input parameters
for the equilibration models were the preshock temperature, the slopes
and normalizations of each powerlaw, shock radius and width.  We
generated 1000 Monte Carlo simulations of these parameters using
the realizations of the density model and drawing preshock temperature
values from a Gaussian distribution based on the mean and uncertainty.
The output equilibration models are the median output model profiles
from this process.


\subsection{Bow shock}

\begin{figure}
  \centering
  \includegraphics[width=0.98\columnwidth]{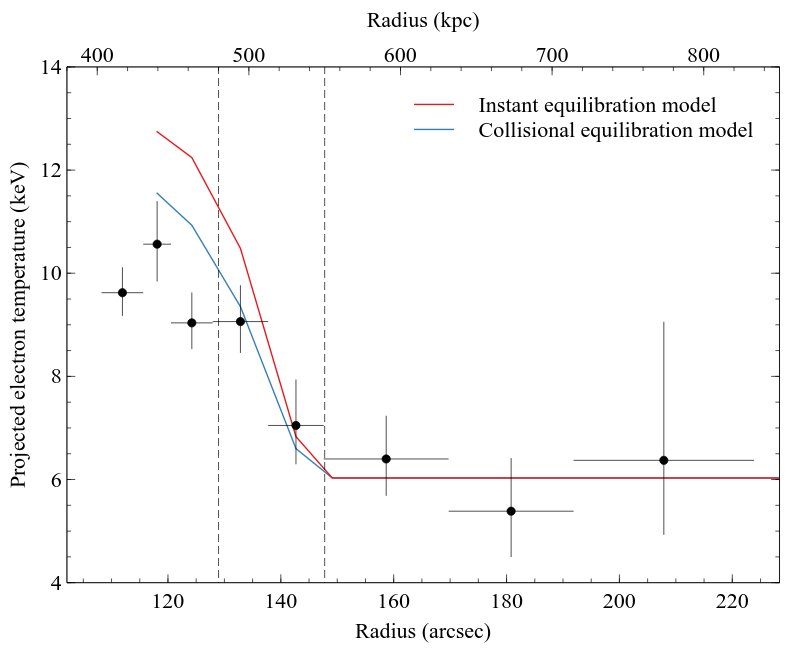}
  \caption{Projected temperature profile extracted in narrow (black points) and broad (white points) radial bins across the bow shock front (from Fig.~\ref{fig:bowtempprofile}) with best-fit models for collisional (blue) and instant equilibration (red) of the electrons and ions behind the shock.}
  \label{fig:boweiplot}
\end{figure}

Fig.~\ref{fig:boweiplot} shows the projected electron temperature
profile across the bow shock and the instant and collisional
equilibration models.  The instant equilibration model is a
particularly poor fit to the observed temperature profile with
$\chi^{2}=46.6$ for 4 degrees of freedom (the postshock datapoints).
We conservatively exclude the datapoint closest to the subcluster's
cool core.  Although not immediately apparent from the images and
temperature map, this region may contain a low level of cooler
stripped material.  In addition, the simplifying assumptions of the
density model breakdown with increasing distance behind the shock (see
section \ref{sec:shockstructure}).  From the p-value of
$2\times10^{-9}$, the observed profile is inconsistent with the
instant equilibration model at the $6\sigma$ level.  A consistent
result is obtained if the preshock temperature is included as a free
parameter rather than assuming the observed value and corresponding
uncertainty.  We therefore rule out the instant equilibration model
for the bow shock in Abell\,2146.

The collisional equilibration model provides an improved fit to the
observed temperature profile with $\chi^{2}=14.0$ for 4 degrees of
freedom.  Although the observed temperature ${>}100\kpc$ behind the
shock appears low compared to the model predictions, this discrepancy
is not particularly significant ($2.5\sigma$).  Adiabatic expansion of
the gas in the postshock region would lower the temperature.  However,
the postshock electron pressure agrees with the prediction from the
shock jump conditions.  The pressure is then constant or increasing
from the shock front to the subcluster's cool core.  Adiabatic losses
do not therefore appear significant over this distance.

Instead, it is likely that the low postshock temperature reflects the changing
state of the preshock gas as the subcluster travels through the core
of the primary cluster.  The equilibration models for the postshock
gas temperature are generated from the shock density jump and the
preshock temperature, which are assumed constant.  However, the
hot atmosphere e.g. $100\kpc$ behind the shock was heated by the
shock ${\sim}10^8\yr$ ago when the preshock conditions were likely
different.  The subcluster is travelling through the primary cluster's
core and the preshock conditions vary depending on the unknown, unperturbed
temperature and density structure.  Therefore, whilst the collisional
equilibration model accurately predicts the immediate postshock
temperature (Fig.~\ref{fig:boweiplot}), we expect increasing
departures from this model with increasing distance behind the shock.
An accurate prediction of the preshock conditions from the path of the
subcluster through the primary is beyond the capabilities of our
existing simple equilibration models and the hydrodynamical
simulations of Abell\,2146.

\subsection{Comparison with other collisionless shocks}


Analyses of cluster merger shock fronts in the Bullet cluster and
Abell\,520 concluded instant equilibration of the electrons and ions
at the $2\sigma$ level (\citealt{Markevitch06}; \citealt{Wang18}).
However, an ALMA and ACA analysis of the Bullet cluster shock by
\citet{DiMascolo19} found that a purely adiabatic electron temperature
change (i.e. no additional electron heating) produced the best agreement
between the thermal Sunyaev-Zeldovich and X-ray results.  Although the
shock properties for these three systems are broadly comparable, it is
possible that they have different structure, which produces different
rates of equilibration.  In future, X-ray microcalorimeters will map
electron and ion temperatures behind nearby shock fronts and increase
the sample size for these equilibration measurements.

Electron heating in low Mach number cluster shocks with high $\beta$
(ratio of thermal to magnetic pressure) has been investigated with
particle-in-cell (PIC) simulations
(e.g. \citealt{Caprioli14,Caprioli14b}; \citealt{Guo14,Guo17,Guo18};
\citealt{Ha18}; \citealt{Xu20}).  \citet{Guo17}, for example, found
additional electron heating over adiabatic compression due to the
growth of whistler waves.  For the bowshock in Abell
2146, with $M=2.24\pm0.09$ and preshock electron temperature
$6.0\pm0.6\keV$, their analytical model predicts this mechanism
would heat the electrons by ${\sim}0.5\keV$ above adiabatic compression.
Unfortunately, the existing data cannot distinguish such a modest
increase in electron heating over the collisional model.

Observations of low Mach number heliospheric shocks have shown
minimal electron heating beyond adiabatic compression
(e.g. \citealt{Thomsen87,Schwartz88,Hull00,Masters11}).  However, the
measured electron distributions show significant deviations from
Maxwellians and they found no conclusive evidence of the dependence of
electron heating on shock parameters
(\citealt{Wilson19,Wilson19b,Wilson20}).  For higher Mach numbers
in supernova remnants, the shock transition is likely turbulent and
disordered because plasma instabilities driven by the ions are
typically reflected back upstream.  Especially in regions where the
shock normal is quasi-parallel to the background magnetic field
(e.g. \citealt{Caprioli15}). Self-consistent kinetic simulations show
that both ions (\citealt{Caprioli14,Caprioli14b}) and
electrons (\citealt{Park15}) are then heated non-adiabatically in the shock
precursor. Evidence of electron collisionless heating in PIC
simulations has been found also for oblique and quasi-perpendicular
shocks (e.g. \citealt{Matsumoto15}; \citealt{Xu20}; \citealt{Bohdan20},
and references therein).

In general, electron heating at collisionless shocks should exhibit a
dependence on the shock velocity and Mach number.  Rapid
equilibration is fostered by the presence of upstream plasma
instabilities driven by accelerated particles.  The bowshock in Abell
2146 appears consistent with this picture with a low Mach number and
minimal electron heating over that produced by adiabatic compression and collisional equilibration.

\subsection{Upstream shock}

\begin{figure}
  \centering
  \includegraphics[width=0.98\columnwidth]{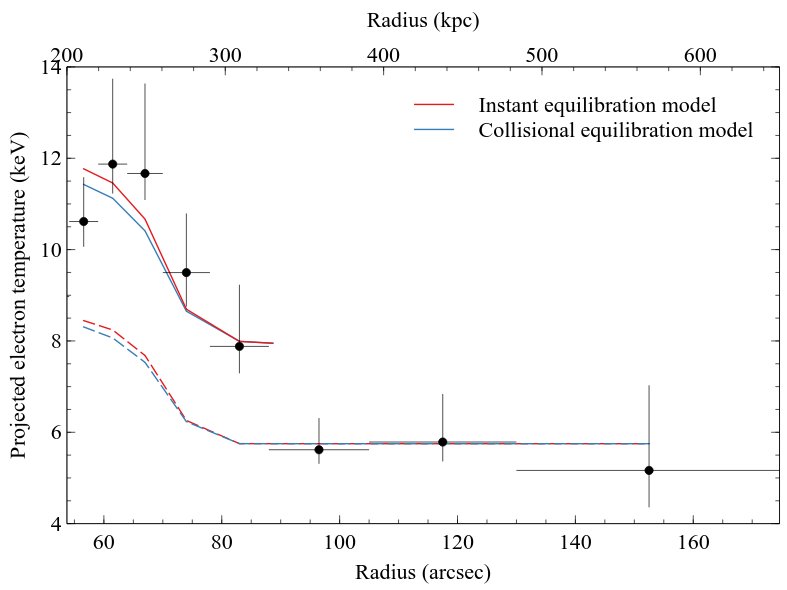}
  \caption{Projected temperature profile extracted in the $30$--$80^{\circ}$ sector across the upstream shock front (from Fig.~\ref{fig:upstreamtempprofile}).  Illustrative models for collisional (blue) and instant (red) equilibration are shown for two different preshock temperatures: $6\keV$ (dashed lines) and $8\keV$ (solid lines).}
  \label{fig:upstreameiplot}
\end{figure}

In principle, the upstream shock front could provide a second test of
electron heating at cluster merger shocks.  However, in practice, the
complex temperature structure at the upstream shock and ram pressure stripping of the primary cluster's galaxy atmospheres downstream prevents an
unambiguous measurement.  Fig.~\ref{fig:upstreamtempprofile} shows the
temperature increasing from $6.0\pm0.6\keV$ at large radius to
$8^{+2}_{-1}\keV$ at a distance of roughly $50\kpc$ ahead of the shock
front.  The projected density jump model shows that this increase
clearly occurs before this narrow shock front.  This temperature
increase could not be identified in our earlier study of this merger.
Based on a preshock temperature of ${\sim}6\keV$, \citet{Russell12}
found that the postshock gas temperature was above the
expectation from the shock jump conditions and the equilibration
models.  Now using a preshock temperature of $8^{+2}_{-1}\keV$ we find
that the postshock temperature is consistent
(Fig.~\ref{fig:upstreameiplot}).  Given the lower Mach number, the
larger uncertainties on the preshock temperature and the complex
temperature structure, the different equilibration models cannot be
distinguished behind the upstream shock.

\section{Systematic uncertainties}

\noindent Here we consider the impact of the key sources of
  systematic uncertainty on the best-fit density jump parameters and
  equilibration analysis for the bow shock.  As discussed in section
  \ref{sec:data}, measured spectral normalization appears on average a
  few per cent higher in the new datasets compared to the 2010
  observations.  This is likely due to the escalating contaminant
  correction.  If the new datasets are `corrected' for this offset to
  match the older observations, the powerlaw normalizations for the
  projected density jump model are systematically lower by $\sim1\%$.
  This uncertainty is insignificant compared to the statistical
  uncertainties.  All other parameters are unaffected.  In addition to
  this calibration systematic, the X-ray spectral modelling makes a
  series of assumptions regarding the applicability of thermal plasma
  emission models (see discussion in section \ref{sec:spec}).  In
  future, X-ray microcalorimeters will start to evaluate the
  systematic uncertainties related to these assumptions.

The new observations were taken over a wide range of roll
  angles so the shock fronts lie on a different part of the ccd in
  each dataset.  The satellite's roll angle measures rotation about
  the viewing axis, which is perpendicular to the ccd or detector plane.  The
  roll angle varies over time to ensure the solar panels can always
  directly view the sun and to maintain preferred operating temperatures.  Observations at different roll angles can be
  advantageous.  It ensures that the results are less dependent on the
  calibration of chip non-uniformities.  However, the shock fronts
  then lie at a range of distances up to $5\amin$ from the ACIS-I
  focal point.  For surface brightness profiles in the $0.5-4\keV$
  band, the majority of the photons have energies below $2\keV$.  At
  these energies, 50 per cent of the photons will fall within a radius
  of $0.5\asec$ for an on-axis source and within $1.5\asec$ at
  $5\amin$ off-axis.  Our analysis assumes an average of $1\asec$ PSF
  blurring, which is appropriate given the large number of
  observations spanning all roll angles.  We tested blurring of
  $0.5\asec$ and $1.5\asec$ and found systematic differences in the
  shock width of $-0.2\kpc$ and $+0.2\kpc$, respectively.  This would
  be a key systematic for much narrower shocks but is not particularly
  significant here.

Galaxy dynamical measurements indicate that the merger axis is
  only $13-19^{\circ}$ from the plane of the sky (\citealt{Canning12};
  \citealt{White15}).  The tightest constraint comes from the small
  line of sight velocity difference between the BCGs, which most
  accurately trace the centres of the respective dark matter halos.
  Whilst hydrodynamical simulations of Abell 2146 indicate an
  inclination angle up to $30^{\circ}$ from the plane of the sky, this
  analysis is based on matching the X-ray morphology and uncertainties
  are expected to be significantly larger (\citealt{Chadayammuri22}).
  The strongest constraint in these simulations is from the standoff
  distance, between the bow shock and the leading edge of the
  subcluster core, which favours inclination angles of
  $15-30^{\circ}$.  \citet{Chadayammuri22} show minimal differences
  between temperature profiles for inclination angles $<20^{\circ}$.
  Inclination effects are therefore expected to be minimal.

We tested the impact of a systematic increase or decrease in the
background by $1\sigma$.  This had a neligible impact on the projected
density model parameters but resulted in a systematic variation in the
measured preshock temperature of $\pm0.3\keV$.  The increase in
cluster surface brightness by roughly an order of magnitude from the
preshock to postshock gas ensured a minimal effect on the measured
postshock temperatures.  The equilibration analysis was repeated with
these systematically shifted temperature profiles and the
corresponding projected density model parameters.  The single fluid
model is ruled out at $5\sigma$ and $7\sigma$ for a systematic
decrease or increase in the background by $1\sigma$, respectively.


The dominant source of systematic uncertainty is the changing state of
the preshock gas as the subcluster travels through the core of the
primary cluster.  The projected density jump and equilibration models
assume a constant shock density jump and preshock temperature.
However, the hot atmosphere at large distances behind the shock was
heated by the shock ${\sim}10^8\yr$ ago, or roughly around core
passage.  The preshock conditions were likely different.  Therefore,
whilst our models should accurately predict the immediate postshock
conditions, the systematic uncertainties increase with distance behind
the shock.  An accurate prediction of the preshock conditions from the
path of the subcluster through the primary is beyond the capabilities
of our existing simple equilibration models and the hydrodynamical
simulations of Abell\,2146.

\section{Conclusions}

The galaxy cluster merger Abell\,2146 features two huge shock fronts
each ${\sim}500\kpc$ in length.  The bow shock, with $M=2.24\pm0.09$,
lies ${\sim}150\kpc$ ahead of the subcluster's core.  The upstream
shock front, $M=1.56\pm0.05$, lies on the far side of the primary
cluster's centre and propagates in the opposite direction.  The new
$2\Ms$ \textit{Chandra} observation of this cluster is the deepest of
merger shock fronts and represents a legacy dataset for studying their
structure.

For the first time, we resolve and measure the width of merger shocks.
The bow shock width is $17\pm1\kpc$ and the upstream shock is
significantly narrower at $10.7\pm0.3\kpc$.  The widths are comparable
to the electron mean free path across each shock at $11\pm2\kpc$ and
$15\pm5\kpc$ for the bow and upstream shocks, respectively.
Collisional shock widths should be a few times the mean free path.
Modest aberrations in the shock shape across the wide sector analysed
further broaden the shock when seen in projection.  These shocks
appear too narrow to be collisional.  Instead, they may be narrow
collisionless shock fronts that appear broader because their smooth
shape is warped by local gas motions.  The widths of both shock fronts
are consistent with local gas motions of $290\pm30\kmps$, which have
modulated the shock speed.  The upstream shock is then expected to be
narrower than the bow shock, as observed, because it forms later in
the merger.  This measurement of turbulence at radii of a few hundred
kpc is consistent with the turbulent level measured in cluster cores
at radii of a few tens of kpc.  This implies a slow increase in
turbulent velocity with radius.

By mapping the electron temperature structure behind the bow shock, we
measure the timescale for the electrons and ions to return to thermal
equilibrium.  For a collisionless shock front, plasma-wave
interactions between the ions and the magnetic field will heat
the massive ions in the narrow shock layer.  The electrons then
subsequently equilibrate with the hotter ions.  We rule out rapid
thermal equilibration, over an unresolved distance, at the $6\sigma$
level.  Instead, the observed electron temperature structure supports
adiabatic compression of the electrons in the shock layer followed by
equilibration on the collisional timescale over ${>}150\kpc$.

Our results for Abell\,2146 are expected to be valid for collisionless shocks with similar parameters in other environments and support the existing picture from the solar wind and supernova remnants.  We rule out strong electron heating but cannot distinguish more modest electron heating over that due to adiabatic compression and collisional equilibration.  The upstream shock is
consistent with these findings but has a more complex structure with a
${\sim}2\keV$ increase in electron temperature ${\sim}50\kpc$ ahead of
the shock's density jump.  Hydrodynamical simulations of Abell\,2146
show this is likely due to a series of additional weaker shocks, which
occur ahead of the upstream shock.  These structures are projected
onto the break up of the primary cluster’s core, which likely obscures
corresponding density increases.

In future, X-ray microcalorimeters will resolve thermal line
broadening and measure both the ion and electron temperatures behind
nearby shock fronts.  With a larger sample, we will determine if the
electron heating rate differs between systems, whether it depends on shock
parameters and search for shock precursors and regions of non-equilibrium ionization.

\begin{table*}
\begin{minipage}{\textwidth}
\caption{}
\begin{center}
\begin{tabular}{l c c c c l c c c c}
\hline
Date & Obs. ID & Aim point & Exposure & Cleaned & Date & Obs. ID & Aim point & Exposure & Cleaned\\
 & & & (ks) & (ks) &  & & & (ks) & (ks) \\
\hline
2010 August 10 & 13020 & I0 & 41.5 & 41.5 &     2019 May 19 & 22225 & I3 & 36.4 & 36.4 \\ 
2010 August 12 & 13021 & I0 & 48.4 & 48.4 &     2019 May 20 & 20915 & I3 & 36.6 & 36.6 \\ 
2010 August 19 & 13023 & I1 & 27.7 & 27.7 &     2019 May 22 & 22228 & I3 & 42.4 & 42.4 \\ 
2010 August 20 & 12247 & I1 & 65.2 & 65.2 &     2019 May 27 & 20919 & I3 & 16.8 & 16.8 \\ 
2010 September 8 & 12245 & I2 & 48.3 & 48.3 &   2019 May 27 & 22235 & I3 & 15.4 & 15.4 \\ 
2010 September 10 & 13120 & I2 & 49.4 & 49.4 &  2019 May 28 & 22236 & I3 & 35.2 & 35.2 \\ 
2010 October 4 & 12246 & I3 & 47.4 & 47.2 &     2019 May 29 & 22237 & I3 & 28.7 & 28.7 \\ 
2010 October 10 & 13138 & I3 & 49.4 & 48.6 &    2019 May 31 & 22238 & I3 & 29.7 & 29.7 \\ 
2018 June 5 & 20555 & I3 & 48.4 & 48.4 &        2019 June 3 & 22030 & I3 & 34.6 & 34.6 \\ 
2018 June 8 & 21101 & I3 & 34.6 & 34.6 &        2019 June 5 & 22241 & I3 & 54.3 & 54.3 \\ 
2018 July 10 & 20556 & I3 & 35.0 & 35.0 &       2019 June 7 & 22242 & I3 & 20.7 & 20.7 \\ 
2018 July 17 & 20557 & I3 & 49.4 & 49.4 &       2019 June 9 & 22243 & I3 & 28.7 & 28.7 \\ 
2018 July 18 & 21130 & I3 & 32.6 & 32.6 &       2019 June 10 & 20560 & I3 & 37.6 & 37.6 \\     
2018 July 24 & 20912 & I3 & 34.2 & 34.2 &       2019 June 13 & 22253 & I3 & 24.7 & 24.7 \\     
2018 July 27 & 21135 & I3 & 21.8 & 21.8 &       2019 June 16 & 22254 & I3 & 19.8 & 19.8 \\     
2018 July 28 & 21136 & I3 & 21.8 & 20.7 &       2019 June 16 & 20562 & I3 & 33.6 & 33.6 \\     
2018 September 7 & 20913 & I3 & 10.8 & 10.8 &   2019 June 19 & 22258 & I3 & 22.8 & 22.8 \\     
2018 September 8 & 21733 & I3 & 56.3 & 56.3 &   2019 June 21 & 22259 & I3 & 19.8 & 19.8 \\     
2018 December 1 & 21970 & I3 & 13.0 & 13.0 &    2019 June 22 & 22260 & I3 & 25.7 & 25.7 \\     
2018 December 2 & 20917 & I3 & 21.8 & 21.8 &    2019 June 24 & 20561 & I3 & 31.6 & 31.6 \\     
2018 December 5 & 20923 & I3 & 17.8 & 17.8 &    2019 June 26 & 22262 & I3 & 13.1 & 13.1 \\     
2018 December 6 & 21996 & I3 & 14.0 & 14.0 &    2019 June 28 & 22263 & I3 & 44.7 & 44.7 \\     
2018 December 15 & 20553 & I3 & 19.8 & 19.8 &   2019 June 29 & 22264 & I3 & 21.8 & 21.8 \\     
2018 December 20 & 22011 & I3 & 19.8 & 19.8 &   2019 June 30 & 22265 & I3 & 32.3 & 32.3 \\     
2018 December 25 & 20914 & I3 & 20.1 & 20.1 &   2019 July 15 & 22200 & I3 & 41.5 & 41.5 \\     
2018 December 26 & 22028 & I3 & 23.8 & 23.8 &   2019 July 22 & 20563 & I3 & 22.8 & 22.8 \\     
2019 January 5 & 22029 & I3 & 32.6 & 32.6 &     2019 July 31 & 20918 & I3 & 22.8 & 22.8 \\     
2019 March 2 & 20921 & I3 & 19.8 & 16.1 &       2019 August 3 & 22678 & I3 & 39.5 & 39.5 \\    
2019 March 2 & 22129 & I3 & 29.7 & 29.7 &       2019 August 5 & 20559 & I3 & 40.1 & 40.1 \\    
2019 March 10 & 22097 & I3 & 33.6 & 30.1 &      2019 August 11 & 22286 & I3 & 16.8 & 16.8 \\   
2019 March 14 & 20554 & I3 & 34.8 & 34.8 &      2019 August 18 & 22093 & I3 & 24.7 & 24.7 \\   
2019 April 6 & 20916 & I3 & 39.5 & 39.5 &       2019 August 20 & 22079 & I3 & 17.6 & 17.6 \\   
2019 April 7 & 20558 & I3 & 27.4 & 27.4 &       2019 August 21 & 22731 & I3 & 39.5 & 39.5 \\   
2019 April 24 & 21674 & I3 & 21.8 & 16.9 &      2019 August 22 & 22732 & I3 & 28.7 & 28.7 \\   
2019 May 9 & 20920 & I3 & 20.9 & 20.9 &         2019 August 23 & 22733 & I3 & 44.5 & 44.5 \\   
2019 May 12 & 22217 & I3 & 17.8 & 17.8 &        2019 August 24 & 22734 & I3 & 34.6 & 34.6 \\   
2019 May 15 & 20552 & I3 & 39.5 & 39.5 &        2019 August 30 & 20922 & I3 & 19.0 & 19.0 \\   
2019 May 17 & 22224 & I3 & 22.8 & 22.8 & & & & & \\ 
\hline
\end{tabular}
\end{center}
\label{tab:obs}
\end{minipage}
\end{table*}

\section*{Acknowledgments}


We thank the reviewer for helpful comments that improved the paper.  HRR acknowledges support from an STFC Ernest Rutherford Fellowship and
an Anne McLaren Fellowship.  Support for this work was provided by the National Aeronautics
and Space Administration through Chandra Award Numbers GO8-19110A, G08-19110B, G08-19110C, G08-19110D and G08-19110E issued by the Chandra X-ray Center, which is operated by the
Smithsonian Astrophysical Observatory for and on behalf of the
National Aeronautics Space Administration under contract NAS8-03060.  This research has made use of data from
the Chandra X-ray Observatory and software provided by the Chandra
X-ray Center (CXC).  We thank Chandra's mission planning team
for this exceptional dataset of such a challenging target.  ARL acknowledges the support of the Gates Cambridge Scholarship, the St John's College Benefactors' Scholarships, NSERC (Natural Sciences and Engineering Research Council of Canada) through the Postgraduate Scholarship-Doctoral Program (PGS D) under grant PGSD3-535124-2019 and FRQNT (Fonds de recherche du Qu\'{e}bec - Nature et technologies) through the FRQNT Graduate Studies Research Scholarship - Doctoral level under grant \#274532.  Many of
the plots in this paper were made using the Veusz software, written by
Jeremy Sanders.

\section*{Data availability}

The \textit{Chandra} data described in this work are available in the \textit{Chandra} data archive (https://cxc.harvard.edu/cda/).  Processed data products detailed in this paper will be made available on reasonable request to the author.

\bibliographystyle{mnras_mwilliams} 
\bibliography{refs.bib}

\clearpage

\end{document}